\documentclass[preprint,prd,aps,showpacs,showkeys,nofootinbib]{revtex4}

\usepackage{amsmath}
\usepackage{amssymb} 
\usepackage{bm}  
\usepackage{bbm}  
\usepackage{graphicx}
\usepackage{hyperref}
\usepackage{float}
\usepackage{dcolumn}  
\usepackage{ragged2e} 
\usepackage{slashed}
\usepackage{booktabs}
\usepackage{siunitx}

\pdfoutput=1 

\usepackage[OT1]{fontenc} 

\graphicspath{{Figures/}}

\begin{document}
	\title{CP-violating multi-field phase transitions and gravitational waves in a hidden NJL sector}
	\author{Chang-Xin Liu}
	\email{liucx@cqu.edu.cn}
	\affiliation{Department of Physics and Chongqing Key Laboratory for Strongly Coupled Physics, Chongqing University, Chongqing 401331, P. R. China}
\begin{abstract}
	We investigate the dynamics of a cosmological first-order phase transition (FOPT) and the associated stochastic gravitational wave background (SGWB) in a hidden strongly coupled sector described by an extended Nambu--Jona-Lasinio (NJL) model with $N_f = 3$ fermion flavors. The model incorporates a CP-violating six-fermion 't Hooft interaction, an explicit chiral symmetry breaking mass term, and chirally symmetric eight-fermion operators that stabilize the vacuum. We perform a multi-field analysis of the tunneling dynamics, going beyond conventional single-field approximations. The interplay between explicit symmetry breaking and CP violation induces a vacuum misalignment, resulting in a curved tunneling path and a spatially varying CP-violating background across the bubble wall. 

    Through a comprehensive scan of the multi-dimensional parameter space, we find a  parameter regime where the conventionally rapid transition rate of the NJL framework is drastically reduced to $\beta/H \sim \mathcal{O}(10^2)$. Consequently, the gravitational wave (GW) production is significantly enhanced, with the predicted SGWB peak amplitudes successfully reaching the detection sensitivity of the proposed $\mu$Ares observatory. Furthermore, our analysis reveals that the macroscopic thermodynamic properties governing the SGWB are predominantly determined by the radial profile of the effective potential, rendering the resulting GW signatures remarkably insensitive to the CP-violating topological vacuum angle. Finally, the explicit symmetry breaking mass introduces a crucial energy bias between competing vacua, triggering the prompt collapse of transient domain wall configurations and thereby ensuring the cosmological viability of the model.
\end{abstract}

	
	\maketitle

\section{Introduction}
\label{sec:intro}
The detection of GWs~\cite{Witten:1984rs,Grojean:2006bp,Caprini:2015zlo} has opened a new observational window on the early Universe, enabling the exploration of cosmological phase transitions at energy scales far beyond the reach of terrestrial experiments. In particular, FOPTs~\cite{Planck:2018vyg,Witten:1984rs,Kibble:1980mv,LISACosmologyWorkingGroup:2022jok} in hidden sectors constitute a well-motivated source of a SGWB. Future space-based interferometers, such as LISA~\cite{LISA:2017pwj,Robson:2018ifk,Babak:2017tow,LISA:2022yao,LISA:2024hlh,Caprini:2015zlo}, Taiji~\cite{Hu:2017mde}, TianQin~\cite{TianQin:2015yph}, BBO~\cite{Corbin:2005ny,Crowder:2005nr},  DECIGO~\cite{Seto:2001qf,Kawamura:2020pcg} and $\mu$Ares\cite{Sesana:2019vho}, are expected to probe a broad class of such signals.

From a theoretical perspective, strongly coupled fermionic sectors exhibiting spontaneous chiral symmetry breaking can be effectively described by NJL-type models~\cite{Nambu:1961tp,Nambu:1961fr,Vogl:1991qt,Wang:2019nhd,Osipov:2014dya,Christian:2025dhe,Buballa:1996tm,Costa:2008gr,Xia:2013caa}. In their minimal realization with only four-fermion interactions, these models typically predict a smooth crossover at finite temperature. However, the inclusion of higher-order multi-fermion operators can qualitatively modify the effective potential and may give rise to a FOPT. Extended NJL-type constructions~\cite{Sakai:2010rp,Kashiwa:2007hw,Fukushima:2003fw,Helmboldt:2019pan,Schwaller:2015tja,Zhao:2026pvt,Tanaka:2026geo} therefore provide a useful effective framework for studying nontrivial vacuum structures, CP violation~\cite{Sakharov:1967dj,Morrissey:2012db,Cohen:1993nk,Cho:1992rv,Crewther:1979pi,Pich:1991fq}, and phase transition dynamics in strongly interacting hidden sectors.

In this context, the six-fermion 't Hooft determinantal interaction is commonly introduced to encode the $U(1)_A$ anomaly and to enhance the strength of the phase transition~\cite{Boer:2008ct,Boomsma:2008gf,Kashiwa:2006rc}. However, this interaction generically destabilizes the effective potential at large field values. This issue can be cured by incorporating chirally symmetric eight-fermion interactions, which stabilize the global vacuum while preserving the barrier structure required for a FOPT~\cite{Osipov:2006ns,Osipov:2006ev,Asakawa:1989bq,Stephanov:1998dy,Allton:2002zi,Stephanov:2004wx,Ratti:2005jh,Ratti:2006gh,Zhang:2006gu,Ghosh:2006qh,Gao:2021nwz}.

An important feature of such extended NJL-type models is their intrinsically multi-field structure. The six-fermion interaction couples scalar and pseudoscalar channels, leading to a nontrivial vacuum manifold. In many existing studies, this structure is simplified by projecting the dynamics onto an effective single-field direction~\cite{Asakawa:1989bq,Helmboldt:2019pan}. While this approximation captures certain thermodynamic aspects, it can miss essential features of the tunneling dynamics, especially in the presence of CP violation.

In this work, we investigate a hidden sector described by an extended NJL-type model with $N_f = 3$ fermion flavors, including a CP-violating six-fermion 't Hooft interaction, an explicit chiral symmetry breaking mass term, and stabilizing eight-fermion interactions. We perform a full multi-field analysis of the finite-temperature phase transition by computing the bounce solution in the complete field space, rather than restricting the dynamics to a single effective direction.

We show that the interplay between explicit symmetry breaking and CP violation induces a misalignment of the vacuum structure in the scalar--pseudoscalar field space. As a result, the tunneling path follows a curved path, and the pseudoscalar condensate develops a nontrivial spatial profile across the bubble wall. This leads to a space-dependent CP-violating background during the transition, providing a qualitatively new dynamical feature compared to single-field treatments.

We further explore the cosmological implications through a parameter scan. While the NJL framework conventionally predicts rapid phase transitions, our results show that a relatively small effective coupling $G\Lambda^2$ can lead to significant supercooling. In this regime, the inverse duration parameter is reduced to $\beta/H \sim \mathcal{O}(10^2) \text{--} \mathcal{O}(10^3)$, which enhances the SGWB to a level detectable by the proposed $\mu$Ares observatory. We also find that the macroscopic thermodynamic properties governing GW generation are determined by the radial profile of the effective potential, making the signal insensitive to the CP-violating phase. Finally, the explicit symmetry breaking term introduces an energy bias between competing vacua, driving the collapse of transient domain wall configurations~\cite{Kibble:1976sj,Zeldovich:1974uw,Saikawa:2017hiv,Hiramatsu:2013qaa,Hiramatsu:2010yz,Gleiser:1998na} and ensuring the cosmological viability of the model.

This paper is organized as follows. In Sec.~\ref{sec2}, we present the theoretical framework and construct the finite-temperature effective potential. In Sec.~\ref{sec3}, we perform the numerical analysis of the phase transition and compute the resulting GW signals. We conclude in Sec.~\ref{sec4}.

\section{The Extended Hidden NJL Model}
\label{sec2}

We consider a strongly coupled hidden sector that is thermally decoupled from the Standard Model thermal bath. The theory contains $N_f = 3$ flavors of Dirac fermions $\chi$, which are singlets under the Standard Model gauge group and carry $N_c = 3$ degrees of freedom associated with a hidden confining interaction \cite{Garcia-Cely:2024ivo}. The underlying gauge dynamics are vector-like, ensuring the absence of gauge anomalies. At low energies, the dominant effective interactions are organized according to an approximate global chiral symmetry $U(3)_L \times U(3)_R$, which emerges in the multi-fermion effective description.

We treat this sector as a generic strongly coupled system capable of undergoing a cosmological FOPT \cite{Kang:2025nhe}. No assumptions are imposed regarding its contribution to the present-day dark matter abundance \cite{Bertone:2004pz,Kribs:2016cew}.

The low-energy dynamics are described by an extended NJL type effective Lagrangian \cite{Osipov:2006ev},
\begin{equation}
	\mathcal{L} = \mathcal{L}_{\text{free}} + \mathcal{L}_{4\chi} + \mathcal{L}_{6\chi}^{\text{CP}} + \mathcal{L}_{8\chi} \, .
\end{equation}

The free part, including explicit chiral symmetry breaking, is given by 
\begin{equation}
	\mathcal{L}_{\text{free}} = \bar{\chi}(i\slashed{\partial} - m_0)\chi \, ,
\end{equation}
where $m_0$ is a small, real, diagonal mass matrix in flavor space. By choosing a real basis for $m_0$, the physical CP-violating phase is transferred to the multi-fermion interaction sector through the combination $\bar{\theta} = \theta_D + \arg\det m_0$. This setup makes the CP-violating phase explicit in the interaction terms.

The four-fermion interaction preserves the approximate chiral symmetry and is written as \cite{Klevansky:1992qe,Hatsuda:1994pi,Vogl:1991qt}
\begin{equation}
	\mathcal{L}_{4\chi} = \frac{G}{2}\sum_{a=0}^{8}
	\Big[(\bar{\chi}\lambda_a\chi)^2 + (\bar{\chi}i\gamma_5\lambda_a\chi)^2\Big],
	\qquad G>0 \, ,
\end{equation}
where $\lambda_a$ are the generators of $U(3)$ with $\lambda_0 = \sqrt{2/3}\,\mathbf{1}_3$.

The six-fermion interaction is given by the ’t Hooft determinantal term \cite{tHooft:1976rip,Kobayashi:1970ji,Kashiwa:2006rc},
\begin{equation}
	\mathcal{L}_{6\chi}^{\text{CP}} = \kappa
	\Big[
	e^{i\theta_D}\det(\bar{\chi}_R \chi_L)
	+
	e^{-i\theta_D}\det(\bar{\chi}_L \chi_R)
	\Big],
	\qquad \kappa < 0 \, ,
\end{equation}
where $\chi_{L,R} = P_{L,R}\chi$ with $P_{L,R} = (1 \mp \gamma_5)/2$, and the determinant is taken over flavor indices. For vanishing phase, $\theta_D = 0$, this reduces to the standard $U_A(1)$-breaking interaction.

The inclusion of the six-fermion interaction can destabilize the effective potential in certain field directions. To ensure boundedness of the potential within the mean-field regime, we include chirally invariant eight-fermion interactions \cite{Osipov:2006ev,Osipov:2006ns,Moreira:2013ura,Hiller:2008rz},
\begin{equation}
	\mathcal{L}_{8\chi} = \frac{g_1}{8}(S_a^2 + P_a^2)^2
	+ \frac{g_2}{8}\Big[
	d_{abc}d_{cde}(S_a S_b - P_a P_b)(S_d S_e - P_d P_e)
	+ 4 f_{ace}f_{bde} S_a S_b P_c P_d
	\Big],
\end{equation}
where $S_a = \bar{\chi}\lambda_a\chi$ and $P_a = \bar{\chi}i\gamma_5\lambda_a\chi$, with $g_1,g_2>0$. These terms represent the most general chirally invariant eight-fermion operators at this order and are assumed to stabilize the mean-field potential for parameters in the physical region.

\subsection{Hubbard--Stratonovich Transformation and Mean-Field Formulation}

To analyze the multi-fermion interactions, we perform a Hubbard--Stratonovich transformation introducing auxiliary scalar and pseudoscalar fields \cite{Eguchi:1976iz,Klevansky:1992qe,Hubbard:1959ub}. After integrating out the fermions, the theory is expressed in terms of bosonic variables within the stationary-phase (mean-field) approximation.

The resulting effective Lagrangian can be written schematically as
\begin{equation}
	\mathcal{L}_{MF} =
	s_a \sigma_a + p_a \phi_a
	+ \frac{G}{2}(s_a^2 + p_a^2)
	+ \mathcal{L}_{6\chi}(s,p)
	+ \mathcal{L}_{8\chi}(s,p) \, ,
\end{equation}
where fermionic determinant contributions are treated separately in the one-loop effective action.

We focus on the flavor-singlet channel, which dominates the phase transition dynamics. Accordingly, we adopt the singlet background approximation \cite{Osipov:2005sp},
\begin{equation}
	\sigma_a = \sigma_0 \delta_{a0}, \qquad \phi_a = \phi_0 \delta_{a0} \, .
\end{equation}

We define the canonically normalized fields via $\lambda_0 = \sqrt{2/3}\,\mathbf{1}_3$ as
\begin{equation}
	\sigma = \sqrt{\frac{2}{3}}\,\sigma_0, \qquad
	\eta = \sqrt{\frac{2}{3}}\,\phi_0 \, .
\end{equation}

After the Hubbard--Stratonovich transformation, the auxiliary fields couple linearly to the fermion bilinears. The fermionic part of the Lagrangian can therefore be written as
\begin{equation}
	\mathcal{L}_{\chi} =
	\bar{\chi}\Big(i\slashed{\partial} - \mathcal{M}\Big)\chi \, .
\end{equation}
This identifies the Dirac operator in the presence of scalar and pseudoscalar background fields. In the mean-field approximation, the background fields $\sigma$ and $\eta$ are proportional to the scalar and pseudoscalar fermion condensates, serving as order parameters for chiral symmetry breaking.

From the Dirac operator above, one identifies the effective mass structure as
\begin{equation}
	\mathcal{M} = m_0 + (\sigma + i\gamma_5 \eta) \, ,
\end{equation}
which contains both scalar and pseudoscalar components.

In this singlet truncation, the $SU(3)$ structure constants simplify significantly. The non-vanishing components involving the singlet index are $d_{0bc} = \sqrt{2/3}\,\delta_{bc}$ and $f_{0bc} = 0$. Consequently, the complicated flavor contractions in the OZI-conserving $g_2$ term become algebraically degenerate with the OZI-violating $g_1$ term. The eight-fermion interaction thus reduces to a single effective coupling $\rho$:
\begin{equation}
	\mathcal{L}_{8\chi} = \frac{\rho}{8}(s_0^2 + p_0^2)^2,
	\qquad \rho \equiv g_1 + \frac{2}{3}g_2 > 0 \, ,
\end{equation}
while the determinantal interaction becomes
\begin{equation}
	\mathcal{L}_{6\chi}^{\text{CP}} =
	\frac{\kappa}{32} A_{000}
	\Big[
	\cos\theta_D\, s_0 (s_0^2 - 3p_0^2)
	- \sin\theta_D\, p_0 (3s_0^2 - p_0^2)
	\Big] \, ,
\end{equation}
where $A_{000} = \frac{2}{3}\sqrt{\frac{2}{3}}$ is a group-theoretic constant arising from the determinant of the $U(3)$ generators in the flavor-singlet direction.

At the mean-field level, the tree-level effective potential $V_{\text{tree}}$ is defined by evaluating the bosonic part of the Lagrangian, $-\mathcal{L}_{MF}$, at the stationary points of the auxiliary fields. Instead of relying on a truncated perturbative expansion in terms of the multi-fermion couplings (which breaks the exact global stability and the Legendre structure of the effective action), we maintain the full non-linear stationary-phase equations. The conditions $\partial \mathcal{L}_{MF} / \partial s_0 = 0$ and $\partial \mathcal{L}_{MF} / \partial p_0 = 0$ yield the exact coupled gap equations connecting the auxiliary condensates $(s_0, p_0)$ to the physical background fields $(\sigma, \eta)$:
\begin{equation}
	\sigma + G s_0 + \frac{\partial \mathcal{L}_{6\chi}^{\text{CP}}}{\partial s_0} + \frac{\partial \mathcal{L}_{8\chi}}{\partial s_0} = 0 \, ,
\end{equation}
\begin{equation}
	\eta + G p_0 + \frac{\partial \mathcal{L}_{6\chi}^{\text{CP}}}{\partial p_0} + \frac{\partial \mathcal{L}_{8\chi}}{\partial p_0} = 0 \, .
\end{equation}

For any given physical field configuration $(\sigma, \eta)$, the auxiliary fields are implicitly determined as $s_0(\sigma, \eta)$ and $p_0(\sigma, \eta)$ by numerically tracking the roots of this highly non-linear system. Substituting these solutions back yields the strictly consistent CJT (Cornwall-Jackiw-Tomboulis) tree-level potential:
\begin{equation}
	V_{\text{tree}}(\sigma, \eta) = - \mathcal{L}_{MF}\Big|_{s_0(\sigma,\eta), p_0(\sigma,\eta)} \, .
\end{equation}

Because the full dynamics of the 't Hooft determinantal interaction and the eight-fermion interaction are retained, the resulting potential intrinsically incorporates the angular dependence induced by the CP-violating phase $\theta_D$ and guarantees global stability without artificial polynomial truncation.

At tree level, the potential does not explicitly depend on the current mass $m_0$; its effects enter through fermionic loop corrections discussed below.

The fermionic determinant depends only on the invariant combination of scalar and pseudoscalar components, leading to an effective constituent mass squared. After integrating out the fermions, the physical mass structure is strictly given by
\begin{equation}
	M^2 = (m_0 + \sigma)^2 + \eta^2 \, .
\end{equation}

The zero-temperature one-loop contribution, regularized using a hard 3D momentum cutoff $\Lambda$, is given by \cite{Coleman:1973jx}
\begin{equation}
	V_{1\text{-loop}}(\sigma, \eta) = -\frac{N_c N_f}{16\pi^2}
	\Big[
	- M^4\ln\left(1+\frac{\Lambda^2}{M^2}\right)
	+ \Lambda^4\ln\left(1+\frac{M^2}{\Lambda^2}\right)
	+ \Lambda^2 M^2
	\Big] \, ,
\end{equation}
defined up to an additive constant fixed by convention.

At finite temperature, the thermal potential for the ideal quasiparticle gas is
\begin{equation}
	V_{\text{thermal}}(\sigma, \eta, T) = -4N_cN_f T \int \frac{d^3p}{(2\pi)^3}
	\ln\left(1 + e^{-E_p/T}\right), \qquad E_p = \sqrt{p^2 + M^2} \, .
\end{equation}

To ensure thermodynamical consistency across the full temperature range up to the high-temperature Stefan-Boltzmann limit, the thermal integral is evaluated over the entire phase space without imposing the hard 3D momentum cutoff $\Lambda$, as the Fermi-Dirac distribution provides a natural and physical ultraviolet suppression.

The full effective potential is then constructed as
\begin{equation}
	V_{\text{eff}}(\sigma, \eta, T) = V_{\text{tree}}(\sigma, \eta) + V_{1\text{-loop}}(\sigma, \eta) + V_{\text{thermal}}(\sigma, \eta, T) \, .
\end{equation}

The temperature dependence is obtained by tracking the global minimum of $V_{\text{eff}}(\sigma,\eta,T)$, which determines the thermodynamic behavior of the phase transition. The exact non-linear multi-field structure derived here forms the robust basis for the bounce analysis and the subsequent calculation of the cosmological phase transition dynamics.

\subsection{Phase Transition Parameters}
\label{sec:PT_parameters}

To quantitatively evaluate the cosmological implications of the FOPT, it is essential to extract the relevant macroscopic parameters. The transition from the metastable symmetric phase to the stable broken phase proceeds via the nucleation of expanding bubbles. At a finite temperature $T$, the thermal decay rate per unit volume is given by \cite{Linde:1981zj}
\begin{align}
	\Gamma(T) \simeq T^4 \bigg(\frac{S_{3}}{2\pi T} \bigg)^{3/2} e^{-S_{3}/T},
\end{align}
where $S_3$ is the three-dimensional Euclidean action. In our CP-violating NJL framework, the effective potential depends on two scalar degrees of freedom, $\sigma$ and $\eta$. To fully capture the dynamic variation of the CP-violating phase, we treat the bubble nucleation as a multi-field bounce problem. The corresponding action is expressed as
\begin{align}
	\frac{S_3}{T} = \frac{4\pi}{T} \int r^{2} \Biggr\{ \frac{1}{2}\left(\frac{d\sigma}{dr}\right)^{2} + \frac{1}{2}\left(\frac{d\eta}{dr}\right)^{2} + V_{eff}(\sigma, \eta, T) \Biggr\} dr,
	\label{eq:S3}
\end{align}
where $r$ is the radial coordinate of the bubble, and the field profiles are obtained by solving the coupled equations of motion numerically \cite{Coleman:1977py}.

The onset of the phase transition is characterized by the nucleation temperature $T_n$. While an approximate criterion (e.g., $S_3/T \approx 140$) is widely used for the electroweak phase transition, it may not be strictly applicable to the distinct energy scales of the hidden NJL sector. To maintain theoretical rigor, we determine $T_n$ by directly evaluating the integral condition \cite{Brdar:2025gyo}:
\begin{equation}
	\int_{T_{n}}^{T_{c}} \frac{dT}{T} \frac{\Gamma(T)}{H(T)^4} = 1.
	\label{eq:Tn}
\end{equation}
This condition implies that, on average, one bubble is nucleated within a single Hubble volume.

Furthermore, to assess the successful completion of the FOPT, we introduce the percolation temperature $T_p$. This temperature corresponds to the epoch when at least 34\% of the comoving volume has transitioned into the true vacuum, which is equivalent to the probability of remaining in the false vacuum being $P_{FV} \approx 0.71$. This probability is defined as \cite{Guth:1981uk,Leitao:2012tx}:
\begin{align}
	P_{FV}(T) &= e^{-I(T)}, \\
	I(T) &= \frac{4\pi v_{w}^3}{3} \int_{T}^{T_{c}} dT^{'} \dfrac{\Gamma(T^{'})}{H(T^{'})T^{'4}} \bigg( \int_{T}^{T^{'}} \frac{dT^{''}}{H(T^{''})} \bigg)^{3}.
\end{align}
In our numerical analysis, both $T_n$ and $T_p$ are rigorously computed through the internal integration routines based on the multi-field bounce solutions. 

For calculating the macroscopic parameters relevant to cosmological observables, we adopt the percolation temperature $T_{\ast} = T_p$ as the characteristic reference point. The inverse duration of the phase transition, normalized by the Hubble parameter, is evaluated as \cite{Caprini:2015zlo}
\begin{equation}
	\frac{\beta}{H_{\ast}} = T_{\ast} \frac{d}{dT}\left(\frac{S_3}{T}\right)\bigg|_{T=T_\ast}\,.
	\label{eq:beta}
\end{equation}

Similarly, the phase transition strength $\alpha$, which quantifies the latent heat released during the vacuum transition relative to the background radiation energy density, is defined as \cite{Caprini:2019egz,Hindmarsh:2020hop}
\begin{align}
	\alpha = \frac{1}{\rho_{\rm rad}(T_{\ast})} \bigg[& \Delta V(\sigma,\eta, T) - T \Delta \left( \frac{\partial V_{eff}(\sigma,\eta, T)}{\partial T} \right) \bigg]_{T=T_{\ast}},
	\label{eq:alpha}
\end{align}
where $\Delta V = V_{false} - V_{true}$. The radiation energy density is given by $\rho_{\rm rad}(T_{\ast})=\pi^2 g_{\ast} T_{\ast}^4/30$. Here, $g_{\ast}$ denotes the effective number of relativistic degrees of freedom in the thermal plasma \cite{Kolb:1990vq,Husdal:2016haj,Borsanyi:2016ksw}, which systematically incorporates contributions from both the Standard Model and the hidden sector at $T_*$. In this work, the publicly available code  \texttt{FindBounce}\cite{Guada:2020xnz} are employed.

\section{Results}
\label{sec3}

\subsection{Numerical Analysis and Vacuum Structure}

To elucidate the structural properties of the effective potential, we first examine the role of the multi-fermion interactions. The 't Hooft determinantal interaction explicitly breaks the $U_A(1)$ symmetry and introduces the CP-violating phase $\theta_D$.

\begin{figure}[htbp] 
	\centering
	\includegraphics[width=0.45\textwidth]{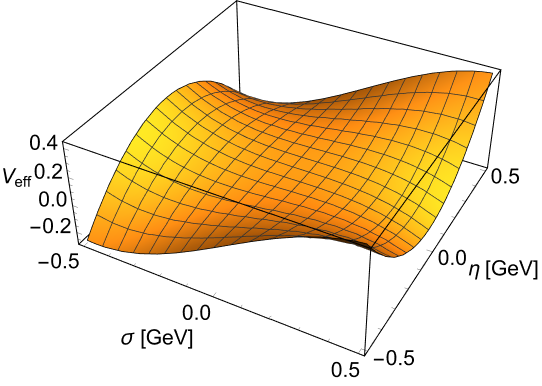}
	\hfill
	\includegraphics[width=0.45\textwidth]{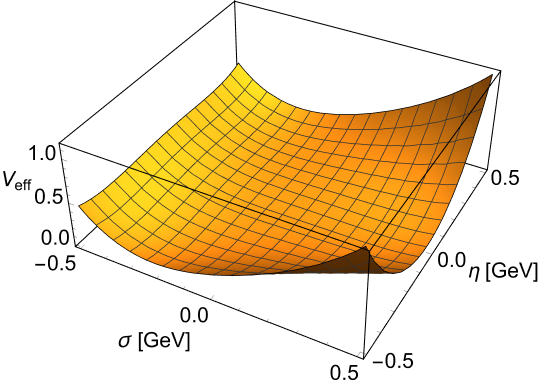} 
	\caption{
		Schematic representation of the effective potential structure in the $(\sigma, \eta)$ plane.
		(Left) Potential including up to six-fermion interactions, exhibiting a topological instability in certain field directions within the mean-field approximation.
		(Right) Globally stabilized potential after the inclusion of eight-fermion interactions, which securely bound the large-field asymptotic behavior.
	}
	\label{fig:6F8F} 
\end{figure}

In the absence of higher-order operators, the six-fermion interaction generates an asymmetric cubic contribution in the radial field direction. As illustrated in the structural schematic in Fig.~\ref{fig:6F8F} (left panel), this term leads to a structural instability where the potential becomes unbounded in certain field directions. This feature reflects the necessity of higher-dimensional operators to ensure a consistent large-field behavior.

The inclusion of chirally invariant eight-fermion interactions modifies the asymptotic behavior of the effective potential. These terms generate quartic contributions that dominate at large field values, ensuring that the potential is globally bounded and stabilized. Consequently, a well-defined physical vacuum emerges, as demonstrated in the structural schematic in Fig.~\ref{fig:6F8F} (right panel).

\begin{figure}[htbp] 
	\centering 
	\includegraphics[width=0.48\textwidth]{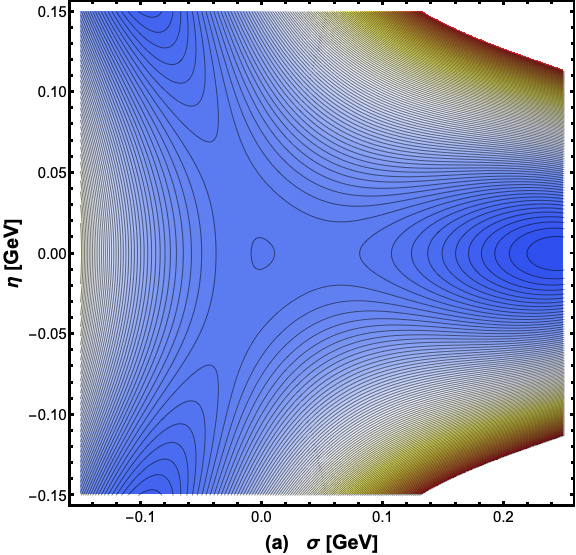}
	\includegraphics[width=0.48\textwidth]{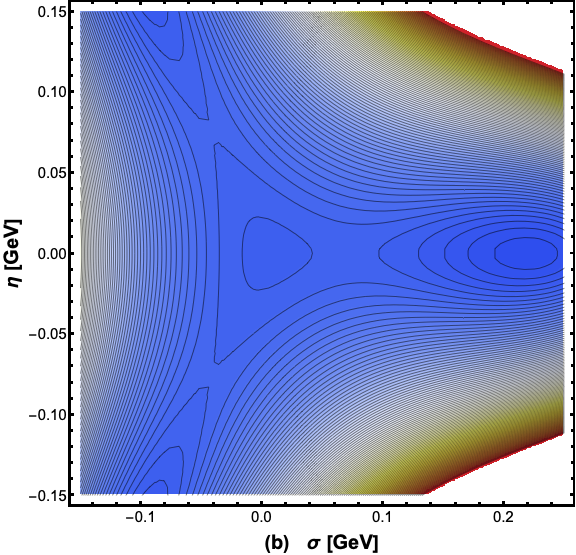}
	\includegraphics[width=0.49\textwidth]{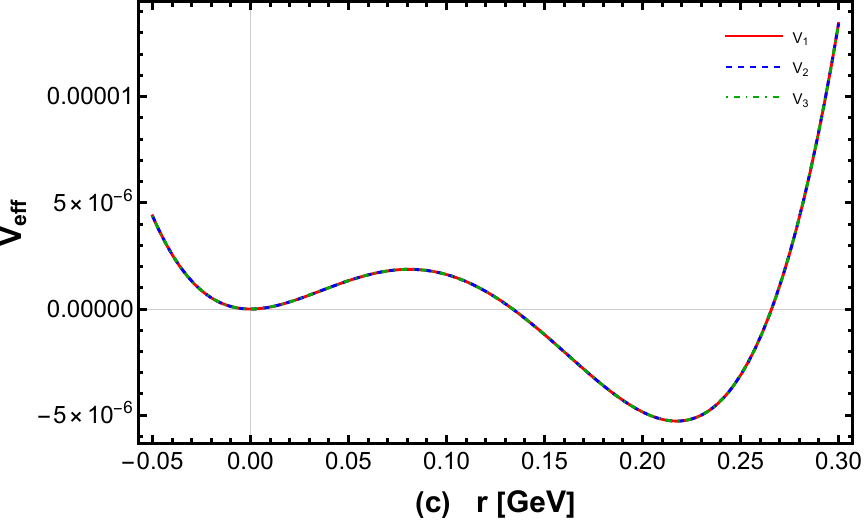} 
	\includegraphics[width=0.49\textwidth]{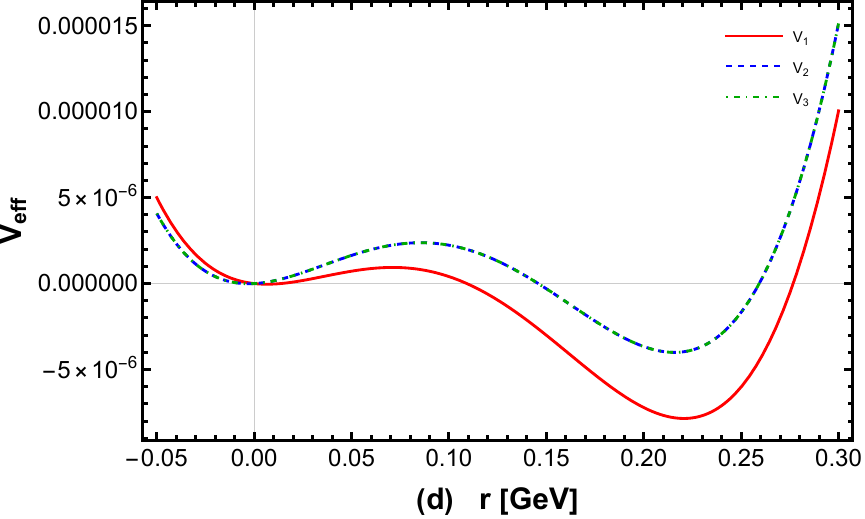} 
	\caption{
		Effect of explicit chiral symmetry breaking on the effective potential.
		(a, c) Chiral limit ($m_0 = 0$), showing three degenerate minima.
		(b, d) Finite current mass case, where the degeneracy is lifted and a unique global minimum emerges.
	}
	\label{fig:currentmass} 
\end{figure}

Next, we examine the effect of explicit chiral symmetry breaking. In the chiral limit ($m_0 = 0$), the potential exhibits three degenerate minima in the $(\sigma,\eta)$ plane (Fig.~\ref{fig:currentmass}a and c), reflecting the residual $Z_3$ structure associated with the six-fermion interaction.

When a finite current mass $m_0$ is introduced, this degeneracy is lifted, leading to a hierarchy among the extrema. As shown in Fig.~\ref{fig:currentmass}b and d, the potential tilt results in a unique configuration becoming the global minimum, while the previously degenerate states become metastable. 

The resulting vacuum structure is governed by the interplay between the CP-violating phase $\theta_D$ and the explicit symmetry breaking induced by $m_0$. As $\theta_D$ is varied over a $2\pi$ interval, the positions of the minima rotate in the $(\sigma,\eta)$ plane due to the underlying $Z_3$ structure. A full return to an equivalent vacuum configuration requires a $6\pi$ evolution in parameter space, reflecting the multi-branch structure of the effective potential.

This structure induces a persistent vacuum misalignment. At tree level, the six-fermion interaction minimizes the potential along angular directions of the form $\phi \sim -\theta_D/3 + 2n\pi/3$, while quantum corrections associated with the explicit mass term tend to align the system along the CP-conserving direction $\phi \sim 0$. The competition between these effects results in a unique global minimum in field space that is generally not radially collinear with the initial symmetric phase. Consequently, the subsequent phase transition must proceed along a curved tunneling path in the multi-field space.

When a finite current mass $m_0$ is introduced, this degeneracy is lifted, leading to a hierarchy among the extrema. As shown in Fig.~\ref{fig:currentmass}b and d, the potential tilt results in a unique configuration becoming the global minimum, while the previously degenerate states become metastable. From a cosmological perspective, lifting this exact $Z_3$ degeneracy is crucial. An explicitly generated energy bias between the true and false vacua creates a volume pressure that drives the collapse of the domain wall network, thereby safely avoiding the cosmological domain wall problem.

\begin{figure}[htbp]
	\centering 
	\includegraphics[width=1\textwidth]{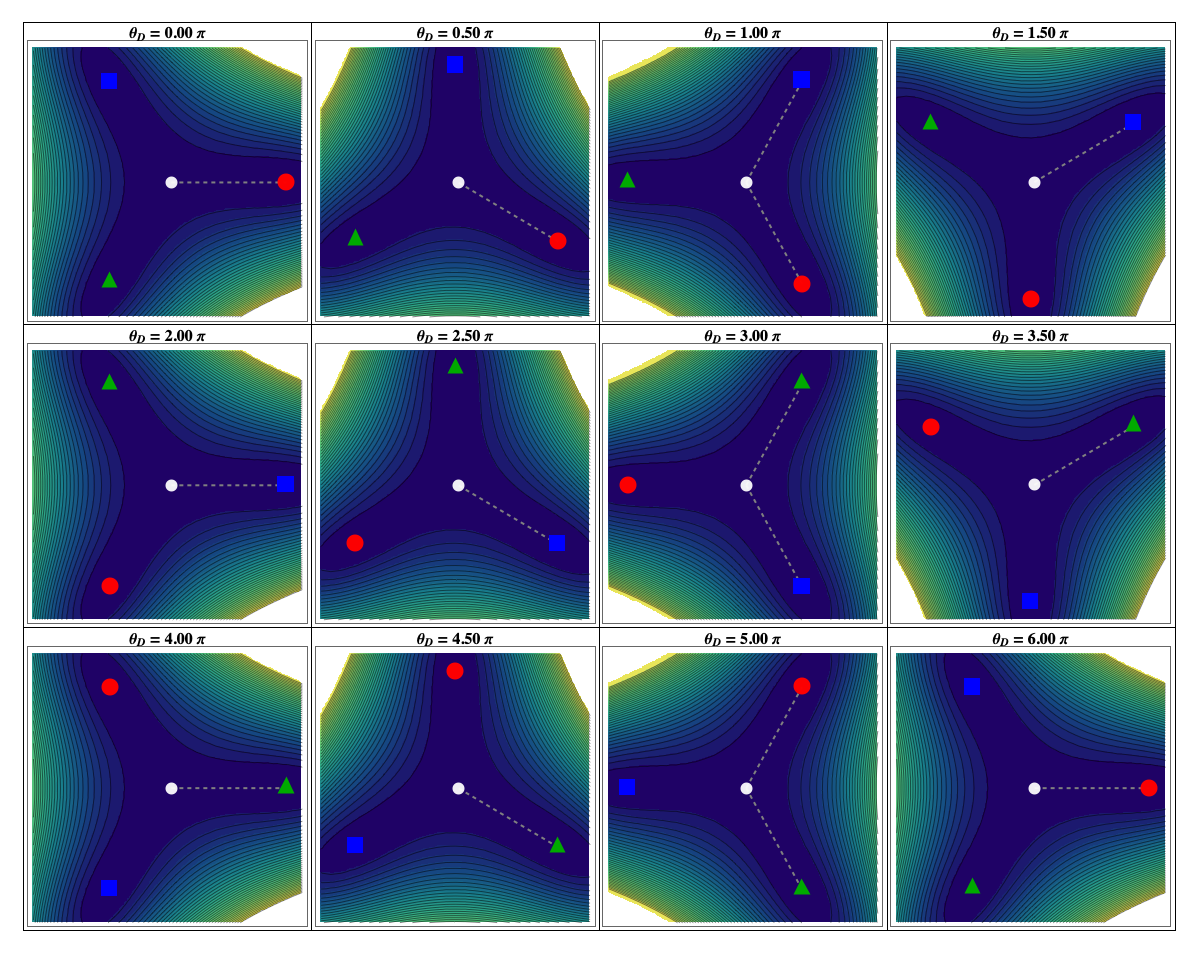}
	\caption{
		Evolution of the effective potential in the $(\sigma, \eta)$ plane as $\theta_D$ varies from $0$ to $6\pi$. 
		The three outer minima rotate continuously, with their azimuthal orientation governed by $\theta_D/3$. 
		The dashed curves represent the curved tunneling trajectories from the central false vacuum (white dot) to the energetically favored true vacua.
	}
	\label{fig:2D-phase}
\end{figure}

Fig.~\ref{fig:2D-phase} illustrates the topological reconfiguration of the effective potential in the $(\sigma, \eta)$ plane as the CP phase $\theta_D$ evolves from $0$ to $6\pi$. For this quantitative evaluation, the effective potential is computed for $N_c = N_f = 3$ at a characteristic temperature $T = 0.2647$. The effective multi-fermion couplings are fixed to dimensionless values $G\Lambda^2 = 8$, $\kappa\Lambda^5 = -200$, and $\rho\Lambda^8 = 100$, defined relative to the cutoff scale $\Lambda = 1$ GeV. A small finite current mass $m_0 = 10^{-4}$ GeV is adopted to explicitly break the chiral symmetry. 

At $\theta_D = 0$, the explicit symmetry breaking induced by $m_0$ lifts the exact $Z_3$ degeneracy, resulting in a unique global true vacuum (red circles) and two higher-energy metastable states (blue squares and green triangles). The dashed lines represent the actual bounce paths from the false vacuum (white dot) to the true vacuum.

As $\theta_D$ increases, these three outer vacua undergo a continuous clockwise rotation in field space. Crucially, while $m_0$ maintains a unique true vacuum at $\theta_D = 0$, the potential restores a twofold degeneracy at specific phase angles $\theta_D = \pi, 3\pi, \text{and } 5\pi$. At these points, two of the minima become energetically equivalent global true vacua, leading to a crossing in the vacuum hierarchy. Notably, while the potential landscape itself is $2\pi$-periodic in $\theta_D$, a full $6\pi$ evolution in parameter space is required for the three labeled vacua to return to their original geometric positions, reflecting the multi-branch topological structure.

To provide a quantitative characterization of this vacuum reconfiguration, we next examine how the energies of the three vacuum states illustrated in Fig.~\ref{fig:2D-phase} evolve as a function of the CP phase.

\begin{figure}[htbp] 
	\centering 
	\includegraphics[width=0.49\textwidth]{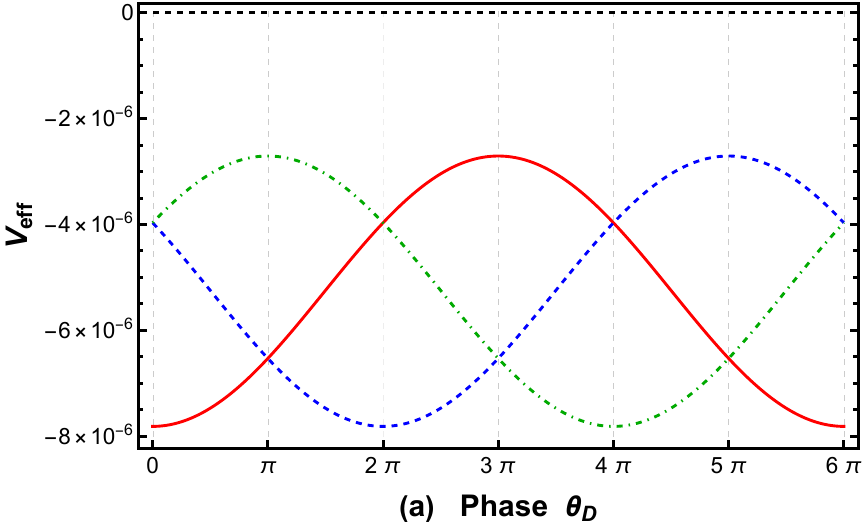}
	\includegraphics[width=0.49\textwidth]{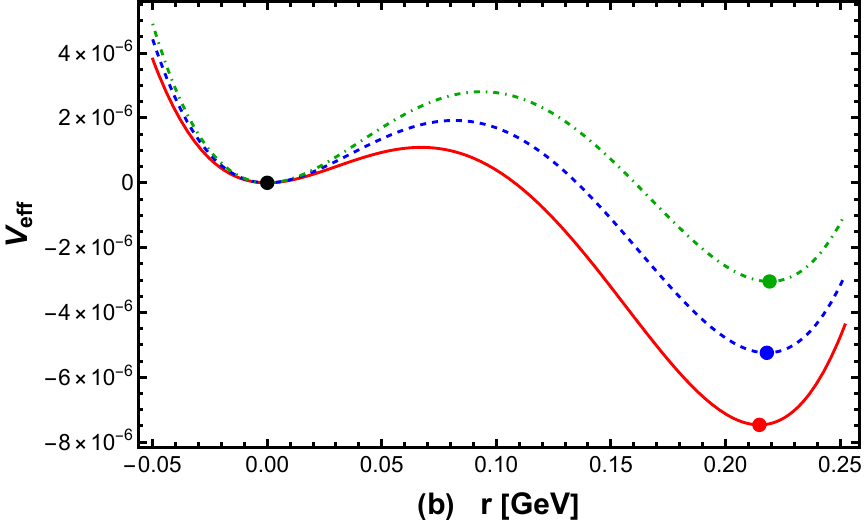}
	\includegraphics[width=0.49\textwidth]{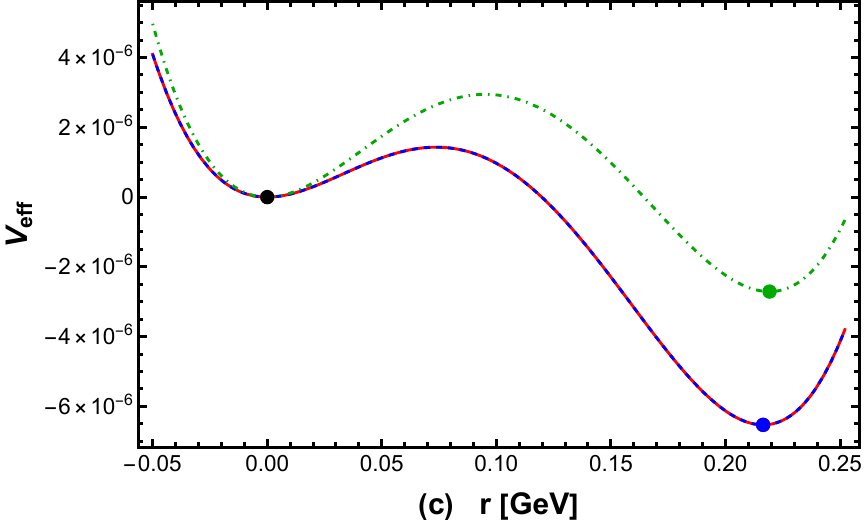} 
	\includegraphics[width=0.49\textwidth]{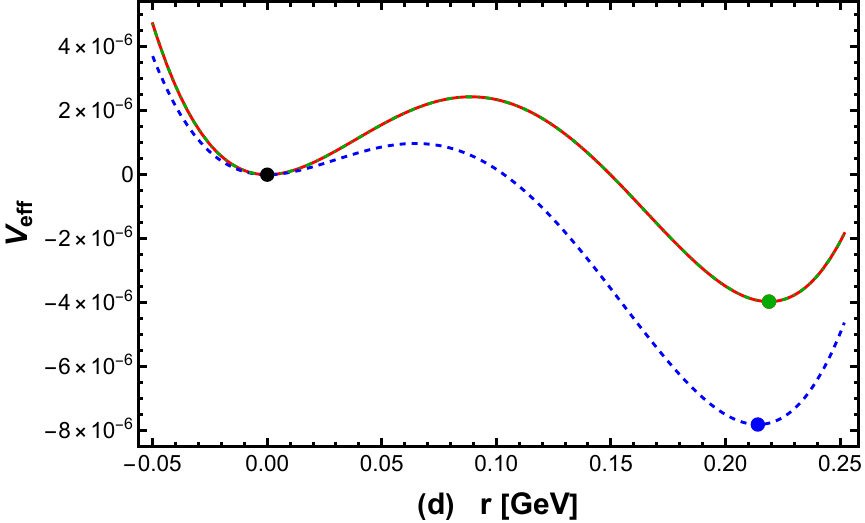} 
	\caption{
		Quantitative evolution of vacuum energies and corresponding cross-sectional profiles. 
		(a) Energy evolution of the three outer minima (red, blue, and green curves, corresponding to the minima shown in Fig.~\ref{fig:2D-phase}) as a function of $\theta_D$. The energy of the central false vacuum is normalized to zero (black dashed line). 
		(b)--(d) One-dimensional cross-sectional profiles of the effective potential at representative values of the CP-violating phase $\theta_D = 0.5\pi, \pi, \text{and } 2\pi$. The ordering of the outer vacua changes with $\theta_D$, reflecting a dynamical reconfiguration of the vacuum hierarchy.
	}
	\label{fig:1Dphase}
\end{figure}

As shown in Fig.~\ref{fig:1Dphase}(a), the energy of the central false vacuum is normalized to zero (black dashed line), while the red, blue, and green curves track the energies of the corresponding outer minima as a function of $\theta_D$. Because these outer minima maintain negative energies relative to the origin, the system invariably favors tunneling away from the symmetric phase. As $\theta_D$ varies, the vacuum energies exhibit a series of intersections, leading to a non-trivial reordering of the local minima.

At each fixed value of $\theta_D$, the lowest-lying state among the outer minima constitutes the global true vacuum, while the others remain metastable. However, the identity of the true vacuum changes continuously as $\theta_D$ evolves, reflecting a dynamical permutation of the vacuum hierarchy driven by the CP-violating phase.

This behavior is further illustrated by the one-dimensional cross-sectional profiles in Fig.~\ref{fig:1Dphase}(b)--(d). At $\theta_D = 0.5\pi$, the degeneracy among the metastable states is completely broken, resulting in a strictly hierarchical structure with a unique global true vacuum and two distinct metastable states at higher energies. At $\theta_D = \pi$, the energy ordering exhibits a level crossing at the global minimum, marked by an exact degeneracy between two outer vacua that coexist as the true vacuum, while the third state resides at a higher energy. At $\theta_D = 2\pi$, a different individual minimum takes over as the unique global true vacuum, accompanied by a restored degeneracy between the remaining two metastable states. This sequence clearly demonstrates the periodic reconfiguration of the vacuum structure and the shifting nature of the true vacuum as a function of the CP-violating phase.

\subsection{Phase Transition Dynamics and Parameter Space}

\begin{figure}[htbp]
	\centering 
	\includegraphics[width=0.49\textwidth]{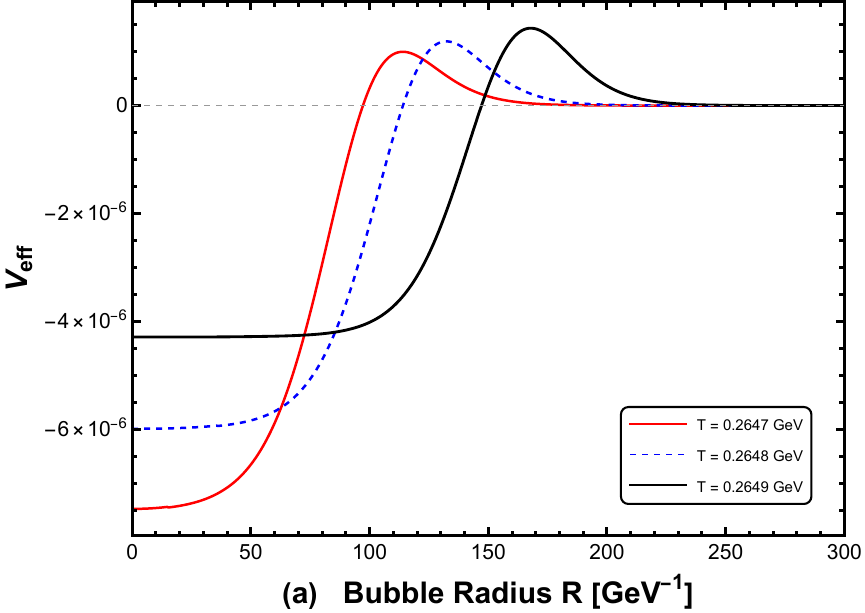}
	\includegraphics[width=0.49\textwidth]{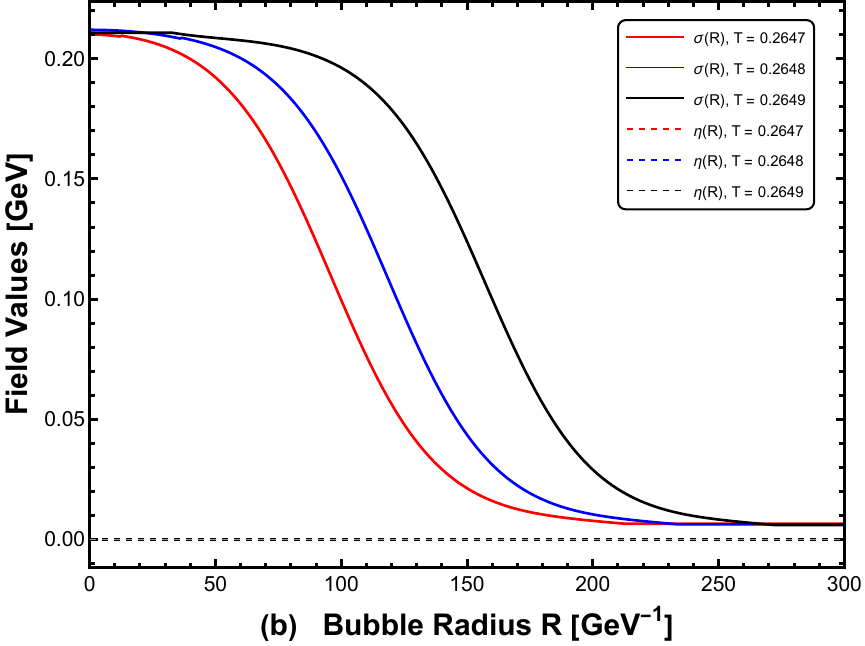}
	\includegraphics[width=0.49\textwidth]{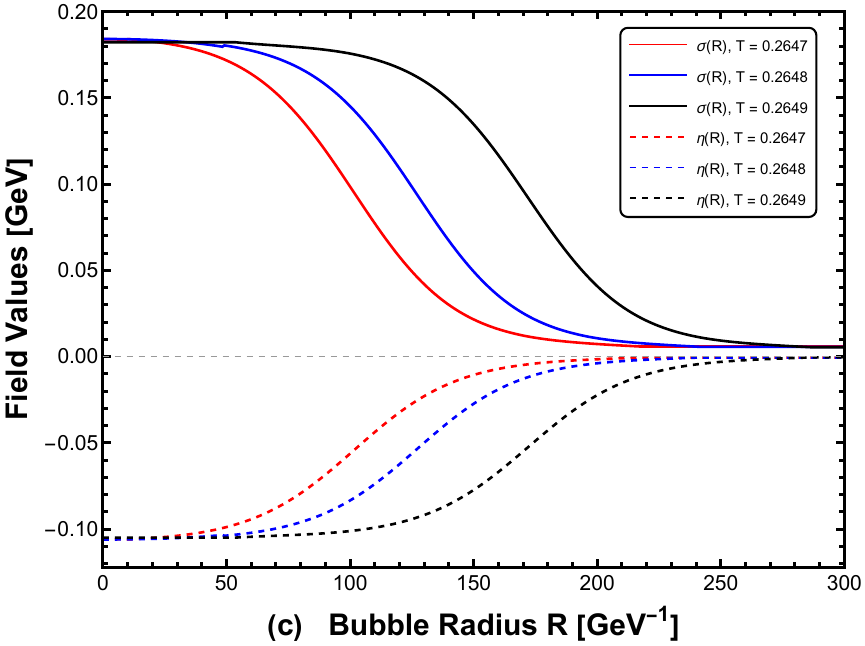}
	\includegraphics[width=0.49\textwidth]{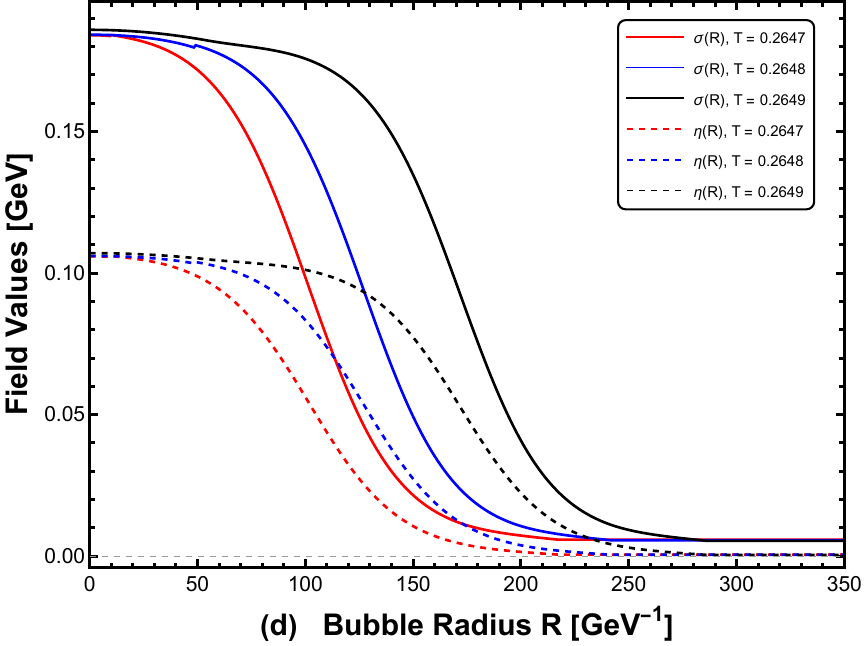}
	\caption{
		Radial bubble profiles and the potential mapping for the NJL model. The fixed parameters are $\Lambda = 1$ GeV, $G\Lambda^2 = 8$, $\kappa\Lambda^5 = -200$, $\rho\Lambda^8 = 100$, and $m_0 = 10^{-4}$ GeV. 
		(a) The effective potential $V_{eff}$ evaluated along the multi-field bounce path $\phi(r) = (\sigma(r), \eta(r))$ for representative temperatures. 
		(b)-(d) The spatial configurations of the scalar field $\sigma(r)$ (solid) and the pseudoscalar field $\eta(r)$ (dashed) for $\theta_D = 0$, $0.5\pi$, and $1.5\pi$, respectively. A non-zero $\theta_D$ forces the bounce path to deviate from the $\sigma$-axis, leading to a localized variation of $\eta(r)$ within the bubble wall.
	}
	\label{fig:profile}
\end{figure}

Having established the vacuum topography, we now examine the structure of the bounce solution for the tunneling process. The phase transition proceeds via the nucleation of critical bubbles, which interpolate between the false vacuum at large radius ($r \to \infty$) and the true vacuum at the bubble center ($r = 0$). 

In Fig.~\ref{fig:profile}(a), we present the effective potential $V_{eff}$ evaluated along the radial bounce path. As the temperature decreases, the deepening of the true vacuum increases the vacuum energy difference $\Delta V$, which enhances the thermodynamic driving force for bubble expansion. This typically leads to a broader field interpolation across the bubble wall, affecting both the critical bubble radius and the associated surface tension.

The multi-field nature of the tunneling is explicitly demonstrated in Figs.~\ref{fig:profile}(b)--(d). 
For $\theta_D = 0$, $0.5\pi$, and $1.5\pi$, both the scalar field $\sigma$ and the pseudoscalar field $\eta$ take non-vanishing values. The spatial configurations of $\sigma$ and $\eta$ evolve continuously with the CP phase, demonstrating that the transition does not reduce to a one-dimensional path. 

For generic $\theta_D$, the vacuum orientation in field space is rotated, forcing the bounce solution to follow a curved path in the $(\sigma, \eta)$ plane. 
As a consequence, the pseudoscalar field develops a nontrivial spatial profile across the thickened bubble wall. In conjunction with the explicit CP-violating phase in the Lagrangian, this dynamic field configuration manifests as a spatially varying CP-violating background, providing the requisite conditions for CP-violating particle transport across the wall.

\begin{figure}[htbp]
	\centering 
	\includegraphics[width=0.49\textwidth]{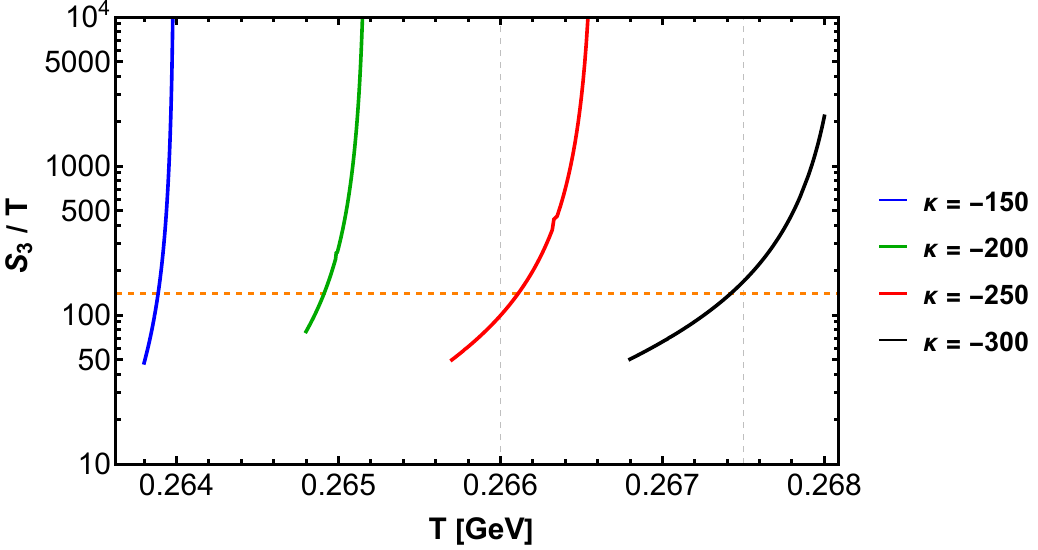}
	\includegraphics[width=0.49\textwidth]{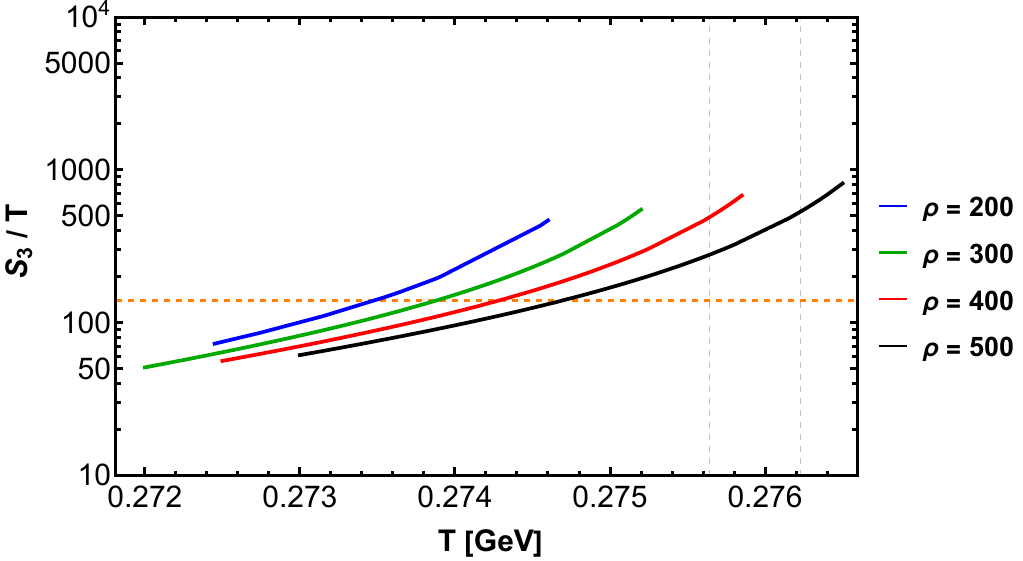}
	\caption{
		The temperature dependence of the normalized three-dimensional Euclidean action $S_3/T$, evaluated along the multi-field bounce paths. The left panel shows the impact of varying the six-fermion coupling $\kappa\Lambda^5$ with a fixed $\rho\Lambda^8 = 100$. The right panel illustrates the effect of varying the eight-fermion coupling $\rho\Lambda^8$ with a fixed $\kappa\Lambda^5 = -500$. The other model parameters are uniformly set to $\Lambda = 1.0\text{ GeV}$, $G\Lambda^2 = 8$, $m_0 = 10^{-4}\text{ GeV}$, and $\theta_D = 0$ for both panels. The horizontal dashed line at $S_3/T = 140$ is shown as a reference value.
	}
	\label{fig:Action-coupling}
\end{figure}

Following the theoretical framework established above, we proceed to numerically evaluate the multi-field bounce solutions and the corresponding Euclidean action $S_3/T$. In Fig.~\ref{fig:Action-coupling}, we present the temperature dependence of $S_3/T$ to systematically analyze the impact of the higher-order multi-fermion interactions, specifically the six-fermion effective coupling $\kappa\Lambda^5$ and the eight-fermion effective coupling $\rho\Lambda^8$. To provide an intuitive visual guide, a horizontal dashed line is drawn at $S_3/T = 140$ as a commonly used reference value, though the actual nucleation temperature $T_n$ is determined by the condition in Eq.~\eqref{eq:Tn}.

The left panel of Fig.~\ref{fig:Action-coupling} illustrates the evolution of $S_3/T$ for various values of $\kappa\Lambda^5$ at a fixed $\rho\Lambda^8$. A larger magnitude of the six-fermion coupling ($|\kappa\Lambda^5|$) results in a higher phase transition temperature. The right panel of Fig.~\ref{fig:Action-coupling} demonstrates the effect of the eight-fermion coupling $\rho\Lambda^8$ at a fixed $\kappa\Lambda^5$. Similar to the effect of the six-fermion interaction, an increase in the coupling $\rho\Lambda^8$ also leads to an increase in the phase transition temperature.

The precise interplay between $\kappa\Lambda^5$ and the stabilizing $\rho\Lambda^8$ thus governs the tunneling action and the subsequent nucleation rate. It provides a rich parameter space to tune the nucleation temperature and the transition duration, which ultimately shape the phase transition parameters $\alpha$ and $\beta/H$.

\begin{figure}[htbp]
	\centering 
	\includegraphics[width=0.7\textwidth]{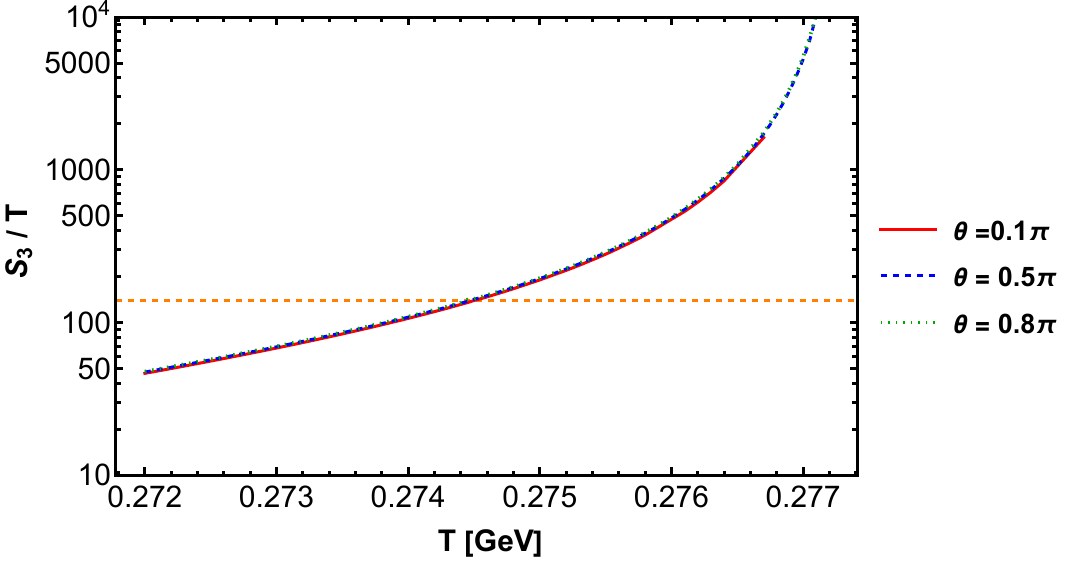}
	\caption{
		The temperature dependence of the normalized tunneling action $S_3/T$ for different CP-violating phases $\theta_D = 0.1\pi$, $0.5\pi$, and $0.8\pi$. The remaining parameters are fixed at $\Lambda = 1.0\text{ GeV}$, $G \Lambda^2= 8$, $\kappa\Lambda^5 = -500$, $\rho\Lambda^8 = 100$, and $m_0 = 10^{-4}\text{ GeV}$.
	}
	\label{fig:Action-theta}
\end{figure}

While the effective couplings $\kappa\Lambda^5$ and $\rho\Lambda^8$ primarily determine the radial structure of the effective potential, it is instructive to examine the dependence on the CP-violating phase $\theta_D$. In Fig.~\ref{fig:Action-theta}, we show the temperature dependence of $S_3/T$ for representative values of $\theta_D = 0.1\pi$, $0.5\pi$, and $0.8\pi$.

The resulting curves for the different CP phases exhibit only minor deviations. As discussed in the vacuum rotation analysis, a non-zero $\theta_D$ primarily induces an angular reorientation of the vacua in field space. While this modifies the specific tunneling path, the radial barrier height and the vacuum energy difference remain predominantly controlled by the effective couplings $G\Lambda^2$, $\kappa\Lambda^5$, and $\rho\Lambda^8$. Consequently, the tunneling action $S_3/T$ is not strictly independent of the CP-violating phase $\theta_D$, but rather  marginal shifts in response to its variations.

We perform a random parameter scan over the dimensionless effective couplings. The scan ranges are set as $G\Lambda^2 \in [3, 25]$, $\kappa\Lambda^5 \in [-1000, -50]$, and $\rho\Lambda^8 \in [50, 1500]$ and $\Lambda=1$ GeV.

The randomly generated parameter sets are subjected to two stringent physical constraints.
First, we impose an effective field theory (EFT) consistency condition, requiring the effective potential to remain smooth and single-valued within the physically accessible field space. Specifically, the auxiliary-field stationary equation may develop non-analytic branch singularities when the pseudoscalar-sector Hessian develops degenerate eigenmodes, corresponding to
\begin{equation}
	\Delta \equiv \left(\frac{3|\kappa|}{16}\right)^2 - 2\rho G > 0 .
\end{equation}
The corresponding singular points in the auxiliary-field space (with mass dimension three) are located at
\begin{equation}
	s_{1,2}
	=
	\frac{\frac{3|\kappa|}{16}\pm \sqrt{\Delta}}{\rho}.
\end{equation}
Parameter sets are rejected if these singularities occur within the EFT validity region,
\begin{equation}
	|s_{1,2}| < \Lambda^3,
\end{equation}
since they would induce unphysical discontinuities in the low-energy effective potential. Conversely, singularities located beyond the cutoff scale are regarded as lying outside the regime of validity of the effective theory and are therefore phenomenologically admissible.

Second, the potential must accommodate a successful FOPT. This requires the existence of a potential barrier separating the false and true vacua, together with the successful completion of bubble nucleation.

\begin{figure}[htbp]
	\centering 
	\includegraphics[width=0.85\textwidth]{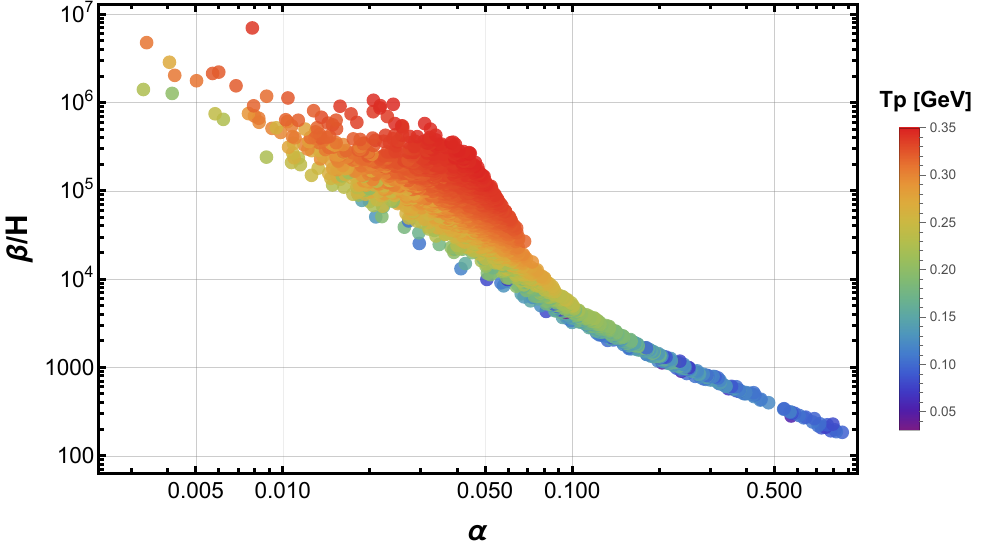}
	\caption{
		Parameter scan results showing the correlation between the phase transition strength $\alpha$ and the dimensionless inverse duration parameter $\beta/H$, where the color mapping represents the percolation temperature $T_p$. 
	}
	\label{fig:A-B}
\end{figure}

The global distribution of the viable parameter space in the $\alpha$--$\beta/H$ plane is shown in Fig.~\ref{fig:A-B}. The plot exhibits a clear correlation between the phase transition strength parameter $\alpha$ and the dimensionless inverse duration parameter $\beta/H$. This behavior can be understood from the underlying vacuum tunneling dynamics: stronger first-order phase transitions are generally associated with enhanced supercooling and delayed bubble nucleation, thereby reducing the inverse duration parameter $\beta/H$.

In particular, parameter regions with relatively large transition strengths, $\alpha \gtrsim \mathcal{O}(10^{-1})$, are predominantly distributed toward smaller values of $\beta/H$, typically in the range $\beta/H \sim 10^2$--$10^3$. Such regions are phenomenologically favorable for the generation of stochastic GWs, since a smaller $\beta/H$ indicates a longer characteristic timescale for the phase transition, thereby strengthening the resulting gravitational-wave signal.

Furthermore, the color mapping in Fig.~\ref{fig:A-B} represents the percolation temperature $T_p$ as an additional physical variable. One observes that the parameter region characterized by large $\alpha$ and small $\beta/H$ is primarily populated by points with comparatively low $T_p$. This trend is consistent with the expected supercooling dynamics: delayed percolation at lower temperatures enhances the vacuum-to-radiation energy ratio, leading to larger values of $\alpha$ and consequently improving the prospects for gravitational-wave detection in future experiments.

\begin{figure}[htbp]
	\centering 
	\includegraphics[width=0.49\textwidth]{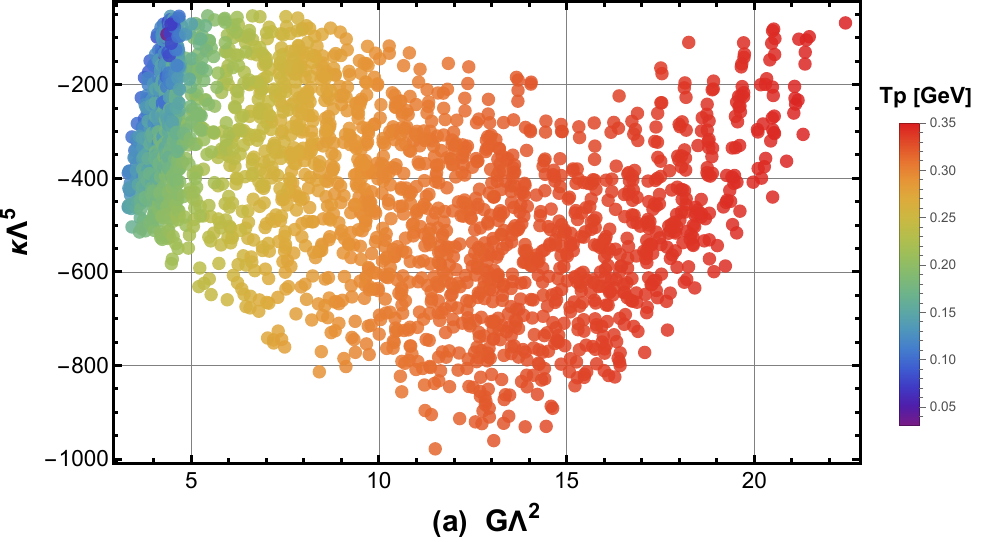}
	\includegraphics[width=0.49\textwidth]{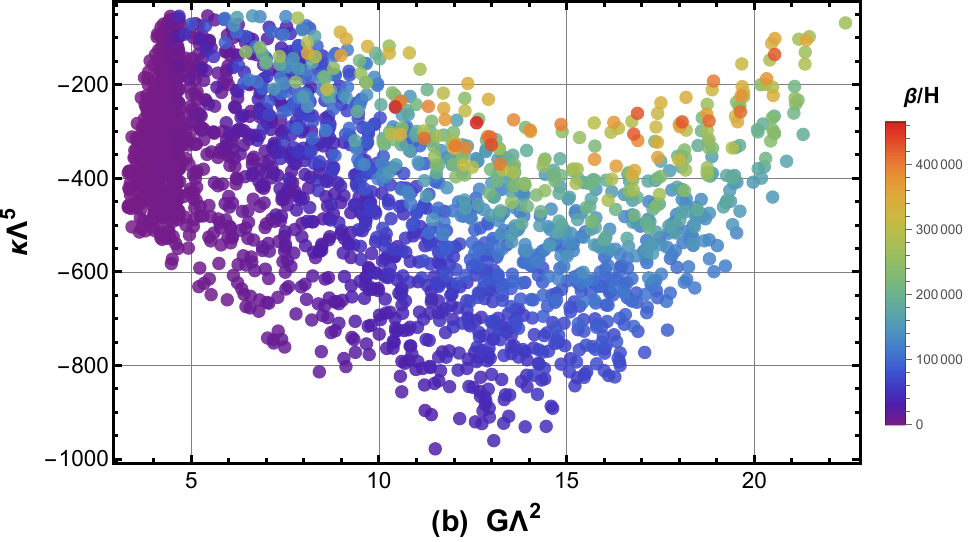}
	\includegraphics[width=0.49\textwidth]{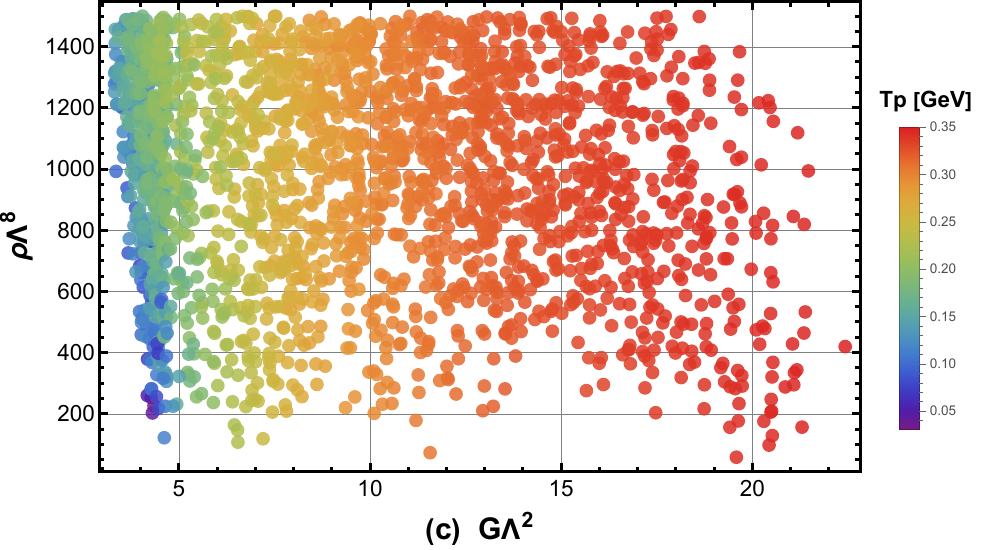}
	\includegraphics[width=0.49\textwidth]{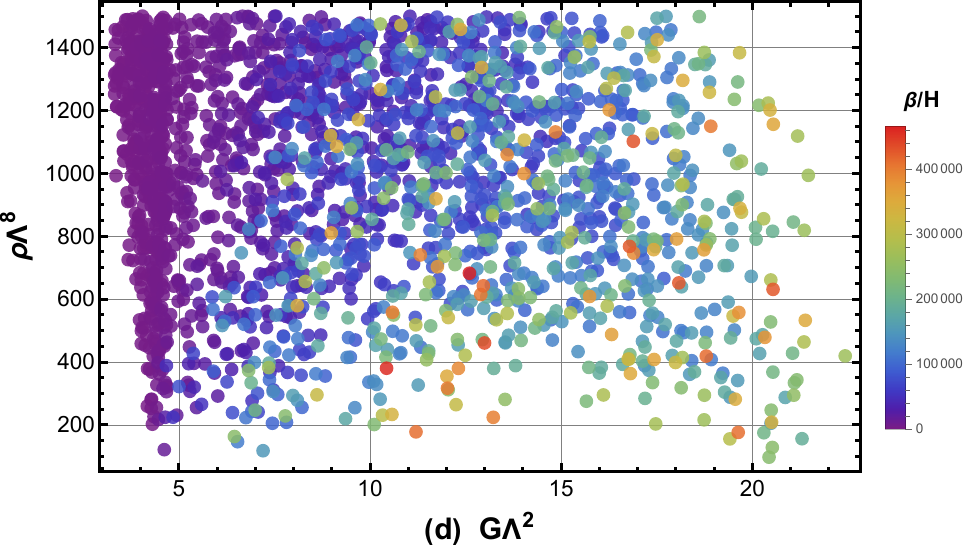}
	\includegraphics[width=0.49\textwidth]{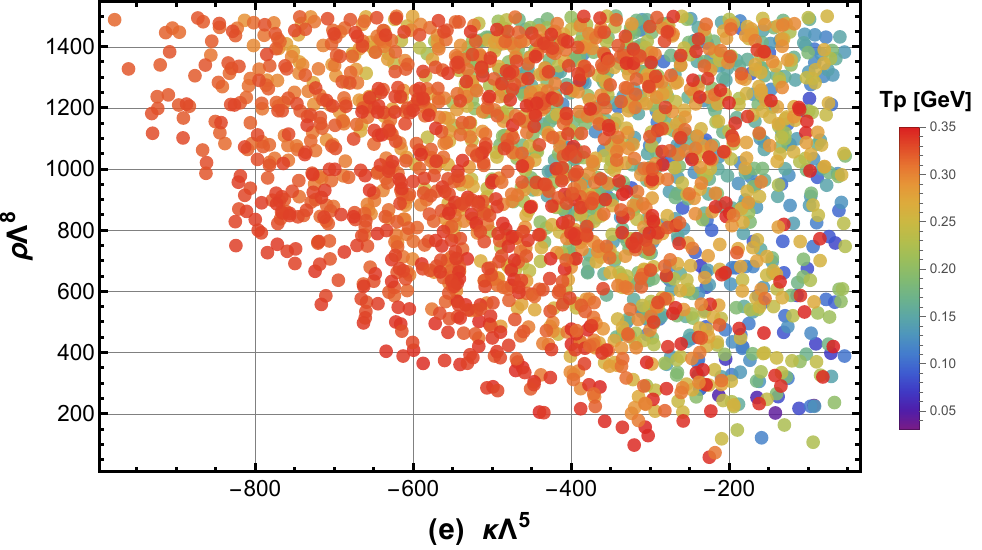}
	\includegraphics[width=0.49\textwidth]{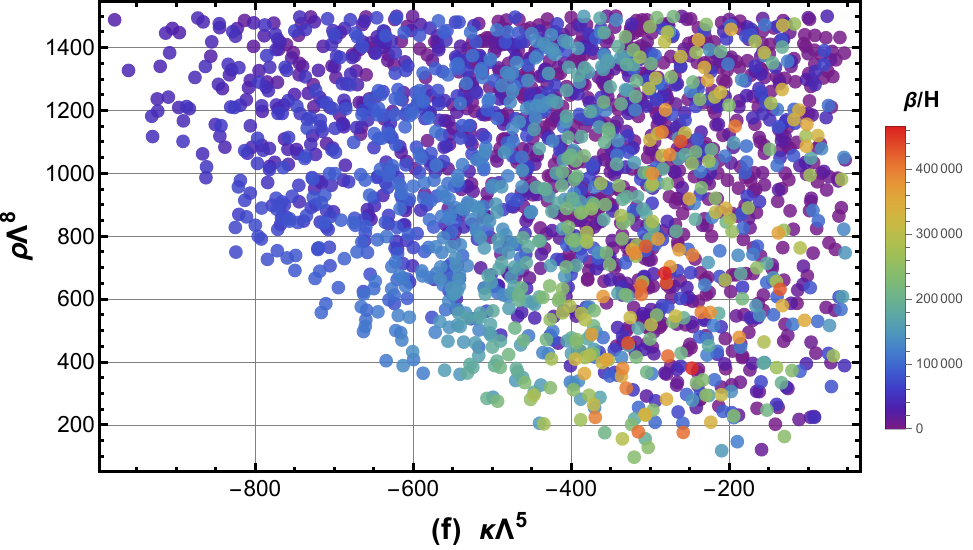}
	\includegraphics[width=0.49\textwidth]{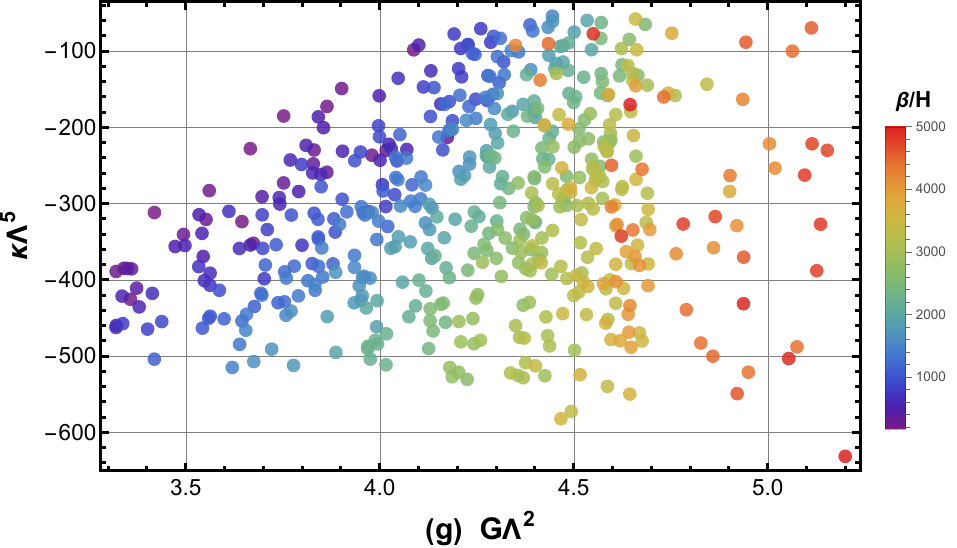}
	\includegraphics[width=0.49\textwidth]{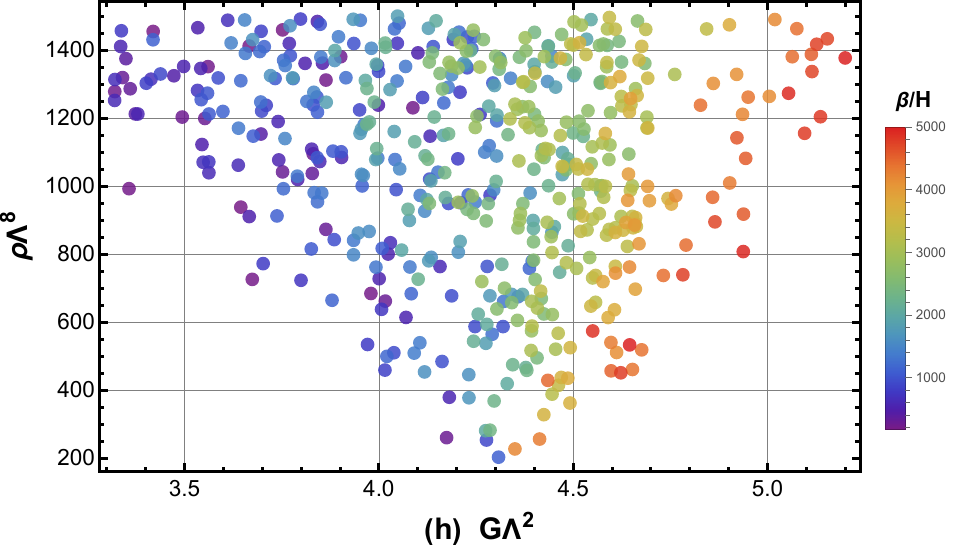}
\caption{Parameter scan results for the parameter space.  Panels (a), (c), and (e) display the correlations among the dimensionless effective couplings ($G\Lambda^2$, $\kappa\Lambda^5$, and $\rho\Lambda^8$), with the color map indicating the percolation temperature $T_p$. Panels (b), (d), and (f) present the corresponding parameter spaces, with the color map indicating the dimensionless inverse duration parameter $\beta/H$. In panels (g) and (h), we specifically select the region with $\beta/H \le 5000$ to further analyze the correlations among the couplings.}
	\label{fig:Scan-coupling}
\end{figure}

The parameter scan results are presented in Fig.~\ref{fig:Scan-coupling}. Panels (a), (c), and (e) show the distributions of the effective couplings, with the percolation temperature $T_p$ represented by the color map. A tendency is observed for lower percolation temperatures to appear in regions characterized by smaller magnitudes of the effective couplings, specifically $G\Lambda^2$, $|\kappa|\Lambda^5$, and $\rho\Lambda^8$. This indicates that the onset of percolation is sensitive to the overall energy scale of the effective interactions.

The corresponding distributions of the inverse duration parameter $\beta/H$ are shown in panels (b), (d), and (f). A similar trend is observed: parameter regions with smaller values of $G\Lambda^2$, $|\kappa|\Lambda^5$, and $\rho\Lambda^8$ generally lead to lower values of $\beta/H$. Since a smaller $\beta/H$ corresponds to a prolonged phase transition duration, these regions are phenomenologically more favorable for the generation of potentially observable stochastic gravitational-wave signals.

Notably, panels (e) and (f) indicate no pronounced direct correlation between $\kappa\Lambda^5$ and $\rho\Lambda^8$. Nevertheless, larger values of $|\kappa|\Lambda^5$ and $\rho\Lambda^8$ tend to be associated with higher percolation temperatures and larger values of $\beta/H$.

To further identify the parameter region relevant for strong phase transition dynamics, panels (g) and (h) restrict the analysis to points satisfying $\beta/H \le 5000$. In this supercooled subset, the viable parameter points are localized in a narrow window of $3 \lesssim G\Lambda^2 \lesssim 5$. This indicates that the four-fermion coupling $G\Lambda^2$ plays a dominant role in determining the gravitational-wave-preferred region of the parameter space. By contrast, the allowed correlations involving $\kappa$ and $\rho$ remain comparatively broad.

\begin{table}[htbp]
	\centering
	\renewcommand{\arraystretch}{1.2} 
	\begin{tabular}{|w{c}{1.5cm}| w{c}{1.6cm} | w{c}{1.8cm} | w{c}{1.8cm} | w{c}{1.6cm} | w{c}{1.6cm} | w{c}{1.6cm} | w{c}{1.4cm} | w{c}{1.6cm}|}
		\hline
		\textbf{BM} & $G/\Lambda^{2}$ & $\kappa/\Lambda^{5}$ & $\rho/\Lambda^8$ & $T_c$ \textbf{[GeV]} & $T_n$ \textbf{[GeV]} & $T_p$ \textbf{[GeV]} & $\alpha$ & $\beta/H$ \\
		\hline
		\hline
		BP1 & 3.7523 & $-185.21$ & 1459.37 & 0.1581 & 0.0945 & 0.0942 & 0.8578 & 183.89 \\ \hline
		BP2 & 3.9019 & $-149.23$ & 1380.43 & 0.1515 & 0.0956 & 0.0949 & 0.6872 & 262.62 \\ \hline
		BP3 & 3.9986 & $-158.73$ & 1238.17 & 0.1494 & 0.1054 & 0.1048 & 0.4283 & 487.33 \\ \hline
		BP4 & 4.2603 & $-70.74$  & 1210.78 & 0.1382 & 0.1045 & 0.1036 & 0.3193 & 758.41 \\ \hline
		BP5 & 4.2787 & $-132.18$ & 763.64  & 0.1175 & 0.0936 & 0.0929 & 0.2462 & 1016.76 \\ \hline
		BP6 & 4.1520 & $-212.83$ & 1040.91 & 0.1601 & 0.1331 & 0.1330 & 0.2024 & 1319.93 \\ \hline
	\end{tabular}
	\caption{Benchmark points for calculating the GW spectra. The critical temperature $T_c$, nucleation temperature $T_n$, and percolation temperature $T_p$ are all presented to illustrate the degree of supercooling.}
	\label{tab:PT_parameters}
\end{table}

\begin{figure}[htbp]
	\centering 
	\includegraphics[width=0.49\textwidth]{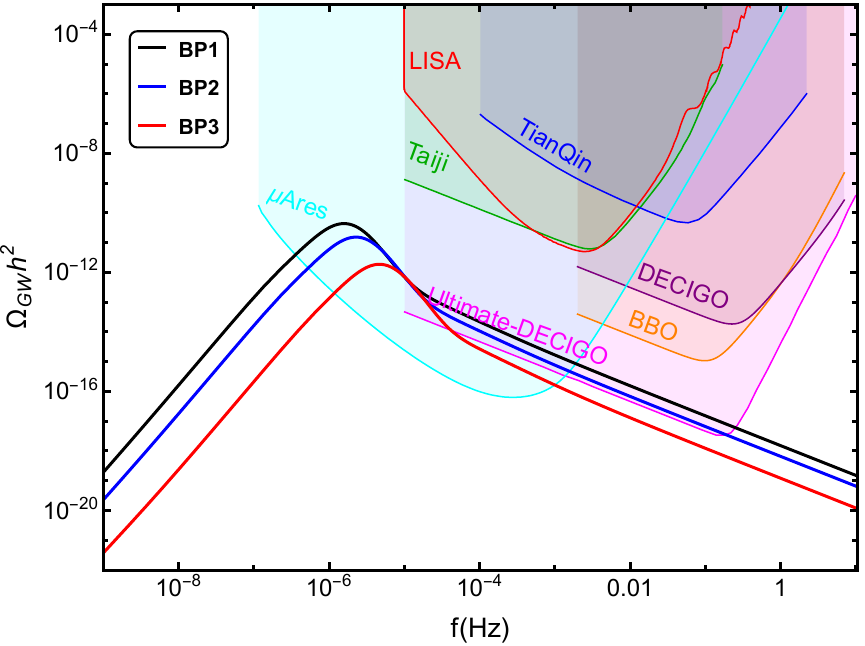}
	\includegraphics[width=0.49\textwidth]{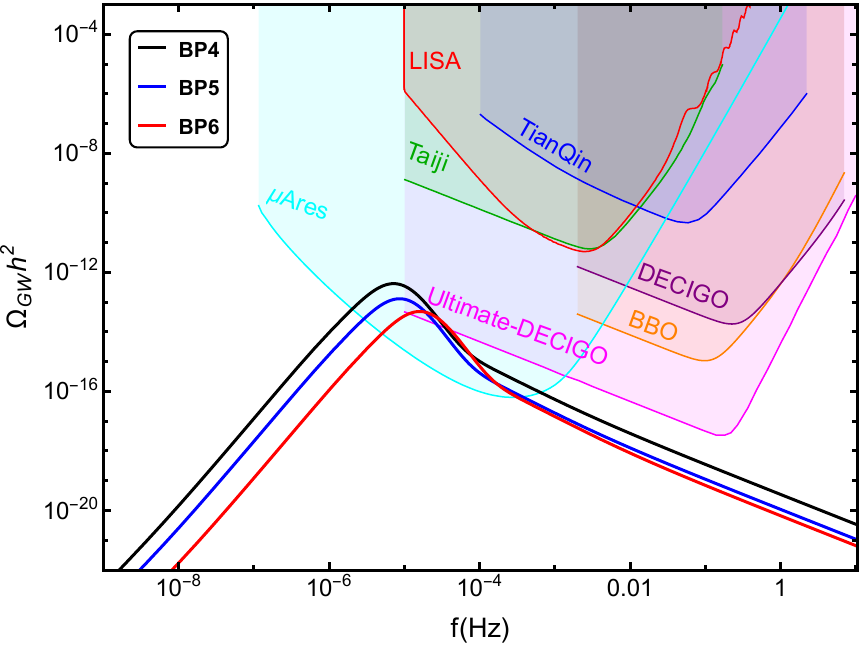}
	\caption{
		Predicted SGWB spectra $\Omega_{\text{GW}}h^2$ generated by the FOPT for the benchmark points BM1--BM3 (left panel) and BM4--BM6 (right panel), plotted alongside the sensitivity curves of future space-based interferometers.
	}
	\label{fig:GW-coupling}
\end{figure}

As illustrated in Fig.~\ref{fig:GW-coupling}, the predicted GW spectra for the selected benchmark points are presented alongside the projected sensitivity curves of future space-based interferometers, including LISA, Taiji, TianQin, BBO, DECIGO, Ultimate-DECIGO, and $\mu$Ares \cite{Sesana:2019vho}. Driven by the supercooling dynamics, the peak frequencies of these GW signals typically fall within the $10^{-6} \sim 10^{-4} \text{ Hz}$ band. The peak amplitudes for all the selected benchmark points are enhanced and lie within the detection sensitivity range of the proposed $\mu$Ares observatory. This suggests that the supercooled phase transitions in our framework may serve as potential observational targets for future multi-messenger astronomy.

Our results indicate that the generation of such detectable GW signals depends sensitively on the four-fermion coupling $G\Lambda^2$. In the specific parameter window of $3 \lesssim G\Lambda^2 \lesssim 5$, the model triggers a sufficiently strong FOPT with a prolonged duration. In this regime, the inverse time scale is reduced to $\beta/H \sim \mathcal{O}(10^2) \text{--} \mathcal{O}(10^3)$. This transition allows the true vacuum bubbles to expand substantially before colliding, thereby increasing the resulting GW energy density.

This enhanced signal generation is confined to a localized region of the parameter space. As the coupling $G\Lambda^2$ increases beyond this window (typically $G\Lambda^2 \gtrsim 5$), the phase transition strength $\alpha$ decreases rapidly, while the inverse duration parameter $\beta/H$ increases to the order of $\mathcal{O}(10^4)\sim \mathcal{O}(10^5)$. Consequently, the phase transition completes more quickly, leading to a suppressed GW energy density (typically dropping below $\Omega_{\text{GW}}h^2 \sim 10^{-18}$), which falls outside the sensitivity range of currently planned detectors.

\begin{figure}[htbp]
	\centering 
	\includegraphics[width=0.47\textwidth]{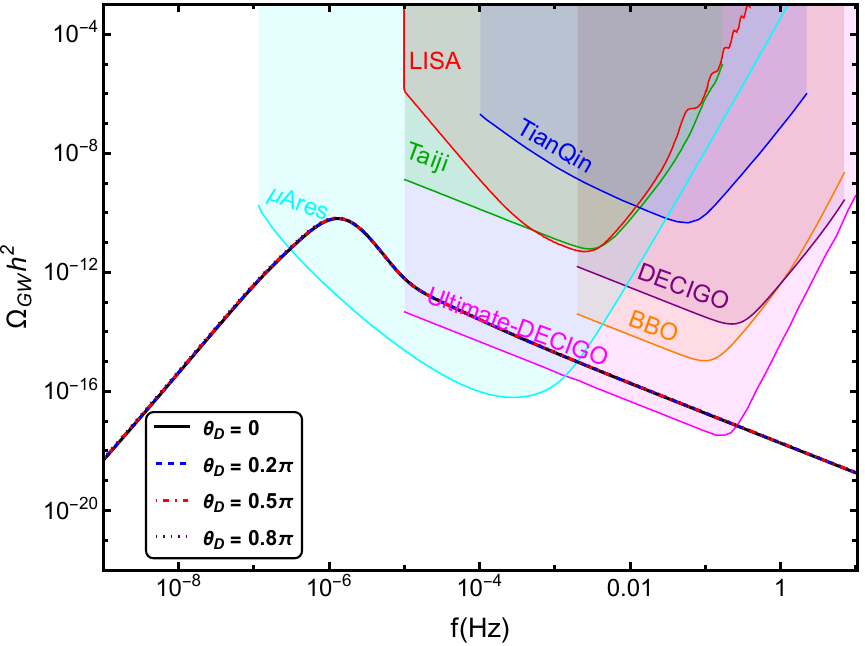}
	\includegraphics[width=0.5\textwidth]{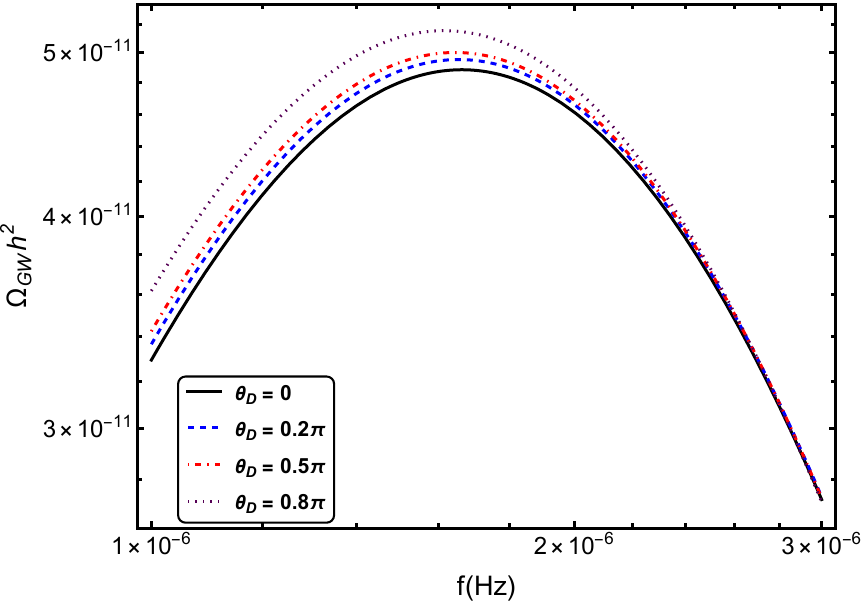}
	\caption{Predicted SGWB spectra for the benchmark point BP1 under varying CP-violating phase angles $\theta_D$. The left panel shows the global spectra plotted against the sensitivities of future detectors, while the right panel provides a zoomed-in view of the peak region, explicitly demonstrating the minimal impact of the phase angle on the gravitational-wave energy density.}
	\label{fig:GW-phase}
\end{figure}

To explicitly evaluate the impact of the CP-violating vacuum structure on the phase transition dynamics, we perform a dedicated analysis using the optimal benchmark point BP1. The generated SGWB spectra under various phase angles $\theta_D$ are presented in Fig.~\ref{fig:GW-phase}. As shown in the left panel, the global spectral profiles across different values of $\theta_D \in [0, 0.8\pi]$ remain highly consistent, with peak amplitudes lying within the detection sensitivity range of the proposed $\mu$Ares observatory. 

To further examine the numerical stability, the right panel of Fig.~\ref{fig:GW-phase} provides a zoomed-in view near the peak frequency band ($f \sim 2 \times 10^{-6}\text{ Hz}$). It is visually evident that variations in the CP phase angle lead to only minor shifts in the GW energy density. This observation is quantitatively supported by the thermodynamic parameters listed in Table~\ref{tab:Phase_Angle}. As the phase angle $\theta_D$ increases, the macroscopic transition parameters undergo only minor variations: the phase transition strength $\alpha$ increases slightly from $0.8578$ to $0.8690$, and the inverse duration $\beta/H$ shifts from $183.89$ down to $179.90$. These stable macroscopic properties confirm that GW production in this framework is primarily determined by the radial profile of the effective potential, and thus the resulting observational signatures are essentially insensitive to the angular vacuum structure parameterized by $\theta_D$.

\begin{table}[htbp]
	\centering
	\renewcommand{\arraystretch}{1.4} 
	\begin{tabular}{|w{c}{1.8cm} | w{c}{2.0cm} | w{c}{2.0cm} | w{c}{2.0cm} | w{c}{1.8cm} | w{c}{1.8cm}|}
		\hline
		$\theta_D/\pi$ & $T_c$ \textbf{[GeV]} & $T_n$ \textbf{[GeV]} & $T_p$ \textbf{[GeV]} & $\alpha$ & $\beta/H$ \\
		\hline
		\hline
		0.0 & 0.1581 & 0.0945 & 0.0942 & 0.8578 & 183.89 \\ \hline
		0.2 & 0.1581 & 0.0944 & 0.0941 & 0.8620 & 182.50 \\ \hline
		0.5 & 0.1581 & 0.0944 & 0.0940 & 0.8640 & 181.90 \\ \hline
		0.8 & 0.1580 & 0.0942 & 0.0934 & 0.8690 & 179.90 \\ \hline
	\end{tabular}
	\caption{Phase transition parameters for the benchmark point BP1 under different CP-violating phase  $\theta_D$.}
	\label{tab:Phase_Angle}
\end{table}

Finally, we briefly comment on the formation and evolution of domain walls. In the chiral limit ($m_0 = 0$), the spontaneous breaking of the approximate $Z_3$ symmetry associated with the six-fermion interaction gives rise to three degenerate vacua. During a cosmological FOPT, different Hubble patches may settle into different vacua, leading to the formation of a network of domain walls. If stable, such a network would eventually dominate the energy density of the Universe and is therefore cosmologically unacceptable.

Introducing an explicit chiral symmetry breaking mass term ($m_0 \neq 0$) effectively lifts this exact $Z_3$ degeneracy. Even in the absence of CP violation ($\theta_D = 0$), the explicit mass term selects a unique global true vacuum and leaves the other equivalent states as higher-energy metastable vacua. The resulting energy difference $\Delta V_{\rm bias}$ generates a volume pressure that drives the collapse of domain walls, preventing the persistence of a stable network. In this way, the explicit symmetry breaking term alone ensures the absence of problematic topological relics.

It is worth noting that we do not compute the GW spectrum generated by the collapse of these domain walls, as it is expected to be subdominant. The dynamics of the wall network are governed by the competition between the volume pressure, induced by the $m_0$-driven energy bias $\Delta V_{\rm bias}$, and the domain wall surface tension. In our framework, the domain wall surface tension is inherently small, as it is set by the relatively low energy scale of the hidden chiral phase transition ($\Lambda = 1.0$ GeV and $T_c \sim \mathcal{O}(0.1)$ GeV). Once the explicit symmetry breaking term lifts the vacuum degeneracy, the resulting volume pressure dominates over this surface tension. As a result, the domain walls are highly unstable and annihilate promptly after bubble percolation, failing to enter a long-lived scaling regime. Any GW radiation emitted during this rapid collapse is expected to be strongly suppressed in amplitude, making it negligible compared to the primary signal generated by the FOPT.

\section{Conclusion}
\label{sec4}

In this work, we have investigated the phase transition dynamics and cosmological implications of a hidden strongly coupled sector within the framework of an extended NJL model. By incorporating a CP-violating six-fermion 't Hooft determinantal interaction together with stabilizing eight-fermion operators, we analyzed the resulting vacuum structure and its impact on the dynamics of cosmological first-order phase transitions.

We find that the explicit chiral symmetry breaking mass $m_0$ lifts the exact $Z_3$ degeneracy of the vacuum, introducing an energy bias among the distinct branches. Meanwhile, the CP-violating phase $\theta_D$ induces a continuous rotation of the vacuum orientation in the scalar--pseudoscalar $(\sigma,\eta)$ field space, leading to a misalignment between competing minima.

This vacuum misalignment modifies the geometry of the tunneling process. In particular, the bounce path is no longer confined to a single field direction but instead follows a curved path in the $(\sigma,\eta)$ plane. As a consequence, the pseudoscalar condensate develops a nontrivial spatial profile across the bubble wall, corresponding to a localized, space-dependent CP-violating background. However, despite this geometric deformation, we find that the impact of $\theta_D$ on the tunneling action remains subdominant.

Through a comprehensive parameter scan, we quantitatively investigated the properties of the phase transition. Although the NJL framework conventionally predicts rapid phase transitions, we find that within a suitable parameter regime—typically with a relatively small effective coupling $G\Lambda^2$—the phase transition duration is significantly prolonged. In this regime, the inverse duration parameter is reduced to $\beta/H \sim \mathcal{O}(10^2) \text{--} \mathcal{O}(10^3)$. Moreover, our results demonstrate that the phase transition strength $\alpha$ and the parameter $\beta/H$ depend primarily on the radial structure of the effective potential, rendering them insensitive to the CP-violating phase $\theta_D$.

We have further computed the SGWB generated by the phase transition. The predicted peak frequencies lie in the band $10^{-6} \sim 10^{-4}\,\mathrm{Hz}$, and the peak amplitudes reach the detection sensitivity of the proposed $\mu$Ares observatory. These results indicate that the extended NJL model is phenomenologically viable, making such hidden sector vacuum dynamics a promising observational target for future space-based gravitational wave astronomy.

Finally, the explicit symmetry breaking induced by $m_0$ introduces a vacuum energy bias that drives the collapse of domain wall networks formed during the transition, thereby helping to ensure the cosmological viability of the model.

Overall, our results demonstrate that while CP violation enriches the vacuum structure and induces nontrivial tunneling trajectories in multi-field space, the macroscopic dynamics of the phase transition and the resulting GW signatures are predominantly controlled by the radial interactions of the effective potential.
	
	\acknowledgments
 This work was supported in part by the Natural Science Foundation of China under Grant No.12447181.

	\appendix
	\begin{onecolumngrid}
	\section{Gravitational wave spectrum}	
		\label{AppGW}
		The GWs power spectra are consisted of three parts, which can be written as\cite{Chen:2017cyc}: \\
		The bubble wall collisions contribution is \cite{Kosowsky:1992vn,Jinno:2016vai,Jinno:2017ixd,Ellis:2019oqb}
		\begin{eqnarray}
			h^2\Omega_\text{co}(f) & = &    1.67 \times 10^{-5}   \, \left(\frac{H_*}{\beta} \right)^{2} \left( \frac{100}{g_*} \right)^{\frac{1}{3}} \left( \frac{\kappa_\eta \alpha}{1+\alpha} \right)^2  
			\left(\frac{0.48\,v_{w}^3}{1+5.3v_{w}^{2}+5v_w^4}\right) \, S_1(f/f_{1}) .
			\label{eq:scadom}
		\end{eqnarray}
		$S_{1}(f/f_{1})$ is the spectral shape, and $f_{1}$ is the peak frequency which are:
		\begin{align}
			S_{1}(r)&=(0.064r^{-3}+0.456r^{-1}+0.48r)^{-1}, \\
			\frac{f_{1}}{ 1\mu \text{Hz}}&=16.5 \left(\frac{\beta}{H_*} \right) \left(\frac{T_*}{100\,{\rm GeV}}\right) \left(\frac{g_*}{100} \right)^{\frac{1}{6}} \left(\frac{f_{*}}{\beta}\right), \\
			\frac{f_{*}}{\beta}&=\frac{0.35}{1+0.069v_{w}+0.69v_w^{4}}.
		\end{align}
		$H_{\ast}$ is the Hubble constant at the temperature $T_{\ast}$, which is selected as the nucleation temperature $T_{n}$ or the percolation temperature $T_{p}$.	Finally, the efficiency factor $\kappa_{\eta}$ determines how much vacuum energy is converted into kinetic energy of the bulk fluid, rather than heating the plasma inside the bubble. This factor can be calculated using the fitting formula \cite{Kamionkowski:1993fg}
		\begin{align}
			\kappa_{\eta}=\frac{1}{1+0.715\alpha}\bigg[ 0.715\alpha+\frac{4}{27}\sqrt{\frac{3\alpha}{2}}\bigg].
		\end{align}
		
		The contribution of sound waves can be calculated using the following fitting formula \cite{Athron:2024xrh,Hindmarsh:2017gnf,Caprini:2019egz,Hindmarsh:2013xza,Hindmarsh:2015qta}
		\begin{align}
			h^2\Omega_\text{sw}(f) =2.061 F_{\text{gw},0} \left(\frac{\kappa_{\text{sw}}\alpha}{1+\alpha}\right)^2 S_{2}(f/f_{2}) \tilde{\Omega}_{\text{gw}} \text{Min}[H_{\ast}R_{\ast}/\bar{U}_{f},1] H_{\ast}R_{\ast},
		\end{align}
		where 
		\begin{align}
			F_{\text{gw},0}&=3.57\times 10^{-5}\bigg(\frac{100}{g_{\ast}}\bigg)^{\frac{1}{3}}, \\
			S_{2}(r)&=r^3\bigg(\dfrac{7}{4+3 r^2}\bigg)^{\frac{7}{2}}, \\
			\frac{f_{2}}{1 \mu \text{Hz}}&= 2.6 \bigg(\frac{z_{p}}{10}\bigg)\bigg(\frac{T_{\ast}}{100 \text{GeV}}\bigg)\bigg(\frac{g_{\ast}}{100}\bigg)^{\frac{1}{6}}\bigg(\frac{1}{H_{\ast}R_{\ast}}\bigg), \\
			\bar{U}_{f}&=\sqrt{\dfrac{3\kappa_{\text{sw}}\alpha}{4(1+\alpha)}}.
		\end{align}
	Here, $z_{p}\sim$ 10 and $\tilde{\Omega}_{\text{gw}}\sim$0.012 are determined from simulations. $R_{\ast}$ is the mean bubble separation and can be roughly calculated by
	\begin{align}
		R_{\ast}=(8\pi)^{1/3}v_{w}/\beta.
	\end{align}
		The suppression factor $\text{Min}[H_{\ast}R_{\ast}/\bar{U}_{f},1]$ accounts the shock formation time effect.
	Based on the bag model and a constant sound velocity $c_{s}=1/\sqrt{3}$, we employ the following fitting formula to estimate the efficiency factor $\kappa_{\text{sw}}$ \cite{Espinosa:2010hh}.
	\begin{align}
		\kappa_{\rm sw} = 
		\begin{cases} 
			\frac{c_s^{11/5}\kappa_A \kappa_B}{\left(c_s^{11/5}  - v_w^{11/5}\right)\kappa_B  + v_wc_s^{6/5}\kappa_A}, & \!\!\!v_w \le  c_s \\[2mm]
			\kappa_B + (v_w-c_s) \delta \kappa + \frac{\left(v_w-c_s\right)^3}{(v_J - c_s)^3}l_{\kappa}, & \!\!\!c_s < v_w < v_J  \\[2mm] 
			\frac{\left(v_J-1\right)^3v_J^{5/2}v_w^{-5/2}\kappa_C \kappa_D}{\left[\left(v_J-1\right)^3 - \left(v_w-1\right)^3\right] v_J^{5/2} \kappa_C + \left(v_w-1\right)^3\kappa_D}, & \!\!\! v_J \le  v_w
		\end{cases}
	\end{align}
	where
	\begin{align}
		\kappa_A &\simeq v_w^{6/5} \frac{6.9  \alpha}{1.36-0.037\sqrt{\alpha} + \alpha},  \\
		\kappa_B  &\simeq \frac{\alpha^{2/5}}{0.017 + \left(0.997 + \alpha\right)^{2/5}}, \\
		\kappa_C &\simeq \frac{\sqrt{\alpha}}{0.135 + \sqrt{0.98 + \alpha}}, \\
		\kappa_D &\simeq \frac{\alpha}{0.73 + 0.083\sqrt{\alpha} + \alpha}, \\
		\delta \kappa &\simeq -0.9{\rm log}\frac{\sqrt{\alpha}}{1 + \sqrt{\alpha}},  \\
		l_{\kappa} &\simeq \kappa_C - \kappa_B - \left(v_J - c_s\right) \delta \kappa, \\
		v_J &= \frac{1}{1+\alpha} \left(c_s + \sqrt{\alpha^2 + \frac{2 \alpha}{3}}\right).
	\end{align}

		The magnetohydrodynamic (MHD) turbulence contribution can be written as \cite{Kahniashvili:2009mf,RoperPol:2019wvy}
		\begin{eqnarray}
			h^2\Omega_\text{turb}(f) &\simeq& 3.35 \times 10^{-4} \, \left(\frac{H_*}{\beta} \right)  \left( \frac{100}{g_*} \right)^{\frac{1}{3}}
			\left(\frac{\kappa_{\rm turb}\,\alpha}{1+\alpha}\right)^{\frac{3}{2}}\,
			v_w \, S_3 (f/f_{3})\,. 
			\label{eq:plasmapresent}
		\end{eqnarray}

		The expressions for $S_{3}$ is:
		\begin{equation} 
	S_{3}(r)=\dfrac{r^3}{(1+r)^{11/3}(1+8\pi f/H_{0})},
		\end{equation}
		and the corresponding peak frequencies is
	\begin{align}
		\frac{f_{turb}}{1 \mu \text{Hz}}= 27 \frac{1}{v_{w}}\bigg(\frac{\beta}{H_{\ast}}\bigg) \bigg(\frac{T_{\ast}}{100 \text{GeV}}\bigg) \bigg(\frac{g_{\ast}}{100}\bigg)^{1/6},
	\end{align}
		the red-shifted Hubble rate at GW generation $H_{0}$ is
		\begin{align}
			H_{0}=16.5\bigg(\frac{g_{\ast}}{100}\bigg)^{1/6}\bigg(\frac{T_{\ast}}{100 \text{GeV}}\bigg) \mu \text{Hz},
		\end{align}
		for the value of $\kappa_{\text{turb}}$, we take $\kappa_{\text{turb}}\approx(5\sim 10)\% \kappa_{\text{sw}}$.
		
	\end{onecolumngrid}	
	
		\bibliography{NJLRef}

@Article{Athron:2024xrh,
  author        = {Athron, Peter and Balazs, Csaba and Fowlie, Andrew and Morris, Lachlan and Searle, William and Xiao, Yang and Zhang, Yang},
  title         = {{PhaseTracer2: from the effective potential to gravitational waves}},
  doi           = {10.1140/epjc/s10052-025-14258-y},
  eprint        = {2412.04881},
  number        = {5},
  pages         = {559},
  volume        = {85},
  archiveprefix = {arXiv},
  journal       = {Eur. Phys. J. C},
  primaryclass  = {astro-ph.CO},
  year          = {2025},
}

@Article{Brdar:2025gyo,
  author        = {Brdar, Vedran and Finetti, Marco and Matteini, Marco and Morais, Ant{\'o}nio P. and Nemev{\v{s}}ek, Miha},
  title         = {{PT2GWFinder: A Package for Cosmological First-Order Phase Transitions and Gravitational Waves}},
  eprint        = {2505.04744},
  archiveprefix = {arXiv},
  month         = {5},
  primaryclass  = {hep-ph},
  year          = {2025},
}

@Article{Chen:2017cyc,
  author        = {Chen, Yidian and Huang, Mei and Yan, Qi-Shu},
  title         = {{Gravitation waves from QCD and electroweak phase transitions}},
  doi           = {10.1007/JHEP05(2018)178},
  eprint        = {1712.03470},
  pages         = {178},
  volume        = {05},
  archiveprefix = {arXiv},
  journal       = {JHEP},
  primaryclass  = {hep-ph},
  year          = {2018},
}

@Article{Vogl:1991qt,
  author       = {Vogl, U. and Weise, W.},
  title        = {{The Nambu and Jona Lasinio model: Its implications for hadrons and nuclei}},
  doi          = {10.1016/0146-6410(91)90005-9},
  pages        = {195--272},
  volume       = {27},
  journal      = {Prog. Part. Nucl. Phys.},
  reportnumber = {TPR-91-6},
  year         = {1991},
}

@Article{Osipov:2014dya,
  author        = {Osipov, A. A. and Hiller, B. and Blin, A. H.},
  title         = {{The 3 Flavor Nambu-Jona-Lasinio with Explicit Symmetry Breaking Interactions: Scalar and Pseudoscalar Spectra and Decays}},
  doi           = {10.5506/APhysPolBSupp.8.183},
  eprint        = {1411.2137},
  number        = {1},
  pages         = {183},
  volume        = {8},
  archiveprefix = {arXiv},
  journal       = {Acta Phys. Polon. Supp.},
  primaryclass  = {hep-ph},
  year          = {2015},
}

@Article{Planck:2018vyg,
  author        = {Aghanim, N. and others},
  title         = {{Planck 2018 results. VI. Cosmological parameters}},
  doi           = {10.1051/0004-6361/201833910},
  eprint        = {1807.06209},
  note          = {[Erratum: Astron.Astrophys. 652, C4 (2021)]},
  pages         = {A6},
  volume        = {641},
  archiveprefix = {arXiv},
  collaboration = {Planck},
  journal       = {Astron. Astrophys.},
  primaryclass  = {astro-ph.CO},
  year          = {2020},
}

@Article{Kibble:1980mv,
  author       = {Kibble, T. W. B.},
  title        = {{Some Implications of a Cosmological Phase Transition}},
  doi          = {10.1016/0370-1573(80)90091-5},
  pages        = {183},
  volume       = {67},
  journal      = {Phys. Rept.},
  reportnumber = {ICTP-79-80-23},
  year         = {1980},
}

@Article{Hindmarsh:2017gnf,
  author        = {Hindmarsh, Mark and Huber, Stephan J. and Rummukainen, Kari and Weir, David J.},
  title         = {{Shape of the acoustic gravitational wave power spectrum from a first order phase transition}},
  doi           = {10.1103/PhysRevD.96.103520},
  eprint        = {1704.05871},
  note          = {[Erratum: Phys.Rev.D 101, 089902 (2020)]},
  number        = {10},
  pages         = {103520},
  volume        = {96},
  archiveprefix = {arXiv},
  journal       = {Phys. Rev. D},
  primaryclass  = {astro-ph.CO},
  reportnumber  = {HIP-2017-02-TH, HIP-2017-02/TH},
  year          = {2017},
}

@Article{Nambu:1961fr,
  author  = {Nambu, Yoichiro and Jona-Lasinio, G.},
  title   = {{Dynamical model of elementary particles based on an analogy with superconductivity. II.}},
  doi     = {10.1103/PhysRev.124.246},
  editor  = {Eguchi, T.},
  pages   = {246--254},
  volume  = {124},
  journal = {Phys. Rev.},
  year    = {1961},
}

@Article{Guth:1981uk,
  author       = {Guth, Alan H. and Weinberg, Erick J.},
  title        = {{Cosmological Consequences of a First Order Phase Transition in the SU(5) Grand Unified Model}},
  doi          = {10.1103/PhysRevD.23.876},
  pages        = {876},
  volume       = {23},
  journal      = {Phys. Rev. D},
  reportnumber = {CU-TP-183},
  year         = {1981},
}

@Article{Leitao:2012tx,
  author        = {Leitao, Leonardo and Megevand, Ariel and Sanchez, Alejandro D.},
  title         = {{Gravitational waves from the electroweak phase transition}},
  doi           = {10.1088/1475-7516/2012/10/024},
  eprint        = {1205.3070},
  pages         = {024},
  volume        = {10},
  archiveprefix = {arXiv},
  journal       = {JCAP},
  primaryclass  = {astro-ph.CO},
  year          = {2012},
}

@Article{Grojean:2006bp,
  author        = {Grojean, Christophe and Servant, Geraldine},
  title         = {{Gravitational Waves from Phase Transitions at the Electroweak Scale and Beyond}},
  doi           = {10.1103/PhysRevD.75.043507},
  eprint        = {hep-ph/0607107},
  pages         = {043507},
  volume        = {75},
  archiveprefix = {arXiv},
  journal       = {Phys. Rev. D},
  reportnumber  = {CERN-PH-TH-2006-125},
  year          = {2007},
}

@Article{Kamionkowski:1993fg,
  author        = {Kamionkowski, Marc and Kosowsky, Arthur and Turner, Michael S.},
  title         = {{Gravitational radiation from first order phase transitions}},
  doi           = {10.1103/PhysRevD.49.2837},
  eprint        = {astro-ph/9310044},
  pages         = {2837--2851},
  volume        = {49},
  archiveprefix = {arXiv},
  journal       = {Phys. Rev. D},
  reportnumber  = {IASSNS-HEP-93-44, FERMILAB-PUB-93-235-A},
  year          = {1994},
}

@Article{Witten:1984rs,
  author       = {Witten, Edward},
  title        = {{Cosmic Separation of Phases}},
  doi          = {10.1103/PhysRevD.30.272},
  pages        = {272--285},
  volume       = {30},
  journal      = {Phys. Rev. D},
  reportnumber = {PRINT-84-0400 (IAS,PRINCETON)},
  year         = {1984},
}

@Article{LISACosmologyWorkingGroup:2022jok,
  author        = {Auclair, Pierre and others},
  title         = {{Cosmology with the Laser Interferometer Space Antenna}},
  doi           = {10.1007/s41114-023-00045-2},
  eprint        = {2204.05434},
  number        = {1},
  pages         = {5},
  volume        = {26},
  archiveprefix = {arXiv},
  collaboration = {LISA Cosmology Working Group},
  journal       = {Living Rev. Rel.},
  primaryclass  = {astro-ph.CO},
  reportnumber  = {LISA CosWG-22-03, FERMILAB-PUB-22-349-SCD},
  year          = {2023},
}

@Article{Linde:1981zj,
  author       = {Linde, Andrei D.},
  title        = {{Decay of the False Vacuum at Finite Temperature}},
  doi          = {10.1016/0550-3213(83)90072-X},
  note         = {[Erratum: Nucl.Phys.B 223, 544 (1983)]},
  pages        = {421},
  volume       = {216},
  journal      = {Nucl. Phys. B},
  reportnumber = {LEBEDEV-81-265},
  year         = {1983},
}

@Article{Caprini:2019egz,
  author        = {Caprini, Chiara and others},
  title         = {{Detecting gravitational waves from cosmological phase transitions with LISA: an update}},
  doi           = {10.1088/1475-7516/2020/03/024},
  eprint        = {1910.13125},
  pages         = {024},
  volume        = {03},
  archiveprefix = {arXiv},
  journal       = {JCAP},
  primaryclass  = {astro-ph.CO},
  reportnumber  = {DESY-19-159, IPPP/19/27, HIP-2019-14/TH, MITP/19-066, IFT-UAM/CSIC-19-139},
  year          = {2020},
}

@Article{Espinosa:2010hh,
  author        = {Espinosa, Jose R. and Konstandin, Thomas and No, Jose M. and Servant, Geraldine},
  title         = {{Energy Budget of Cosmological First-order Phase Transitions}},
  doi           = {10.1088/1475-7516/2010/06/028},
  eprint        = {1004.4187},
  pages         = {028},
  volume        = {06},
  archiveprefix = {arXiv},
  journal       = {JCAP},
  primaryclass  = {hep-ph},
  reportnumber  = {CERN-PH-TH-2010-027},
  year          = {2010},
}

@Article{Wang:2019nhd,
  author        = {Wang, Lingxiao and Jiang, Yin and He, Lianyi and Zhuang, Pengfei},
  title         = {{Chiral vortices and pseudoscalar condensation due to rotation}},
  doi           = {10.1103/PhysRevD.100.114009},
  eprint        = {1901.04697},
  number        = {11},
  pages         = {114009},
  volume        = {100},
  archiveprefix = {arXiv},
  journal       = {Phys. Rev. D},
  primaryclass  = {nucl-th},
  year          = {2019},
}

@article{Caprini:2015zlo,
    author = "Caprini, Chiara and others",
    title = "{Science with the space-based interferometer eLISA. II: Gravitational waves from cosmological phase transitions}",
    eprint = "1512.06239",
    archivePrefix = "arXiv",
    primaryClass = "astro-ph.CO",
    reportNumber = "DESY-15-246",
    doi = "10.1088/1475-7516/2016/04/001",
    journal = "JCAP",
    volume = "04",
    pages = "001",
    year = "2016"
}

@article{LISA:2017pwj,
    author = "Amaro-Seoane, Pau and others",
    collaboration = "LISA",
    title = "{Laser Interferometer Space Antenna}",
    eprint = "1702.00786",
    archivePrefix = "arXiv",
    primaryClass = "astro-ph.IM",
    month = "2",
    year = "2017"
}

@article{Robson:2018ifk,
    author = "Robson, Travis and Cornish, Neil J. and Liu, Chang",
    title = "{The construction and use of LISA sensitivity curves}",
    eprint = "1803.01944",
    archivePrefix = "arXiv",
    primaryClass = "astro-ph.HE",
    doi = "10.1088/1361-6382/ab1101",
    journal = "Class. Quant. Grav.",
    volume = "36",
    number = "10",
    pages = "105011",
    year = "2019"
}

@article{Babak:2017tow,
    author = "Babak, Stanislav and Gair, Jonathan and Sesana, Alberto and Barausse, Enrico and Sopuerta, Carlos F. and Berry, Christopher P. L. and Berti, Emanuele and Amaro-Seoane, Pau and Petiteau, Antoine and Klein, Antoine",
    title = "{Science with the space-based interferometer LISA. V: Extreme mass-ratio inspirals}",
    eprint = "1703.09722",
    archivePrefix = "arXiv",
    primaryClass = "gr-qc",
    doi = "10.1103/PhysRevD.95.103012",
    journal = "Phys. Rev. D",
    volume = "95",
    number = "10",
    pages = "103012",
    year = "2017"
}

@article{LISA:2022yao,
    author = "Seoane, Pau Amaro and others",
    collaboration = "LISA",
    title = "{Astrophysics with the Laser Interferometer Space Antenna}",
    eprint = "2203.06016",
    archivePrefix = "arXiv",
    primaryClass = "gr-qc",
    doi = "10.1007/s41114-022-00041-y",
    journal = "Living Rev. Rel.",
    volume = "26",
    number = "1",
    pages = "2",
    year = "2023"
}

@article{LISA:2024hlh,
    author = "Colpi, Monica and others",
    collaboration = "LISA",
    title = "{LISA Definition Study Report}",
    eprint = "2402.07571",
    archivePrefix = "arXiv",
    primaryClass = "astro-ph.CO",
    month = "2",
    year = "2024"
}

@article{TianQin:2015yph,
    author = "Luo, Jun and others",
    collaboration = "TianQin",
    title = "{TianQin: a space-borne gravitational wave detector}",
    eprint = "1512.02076",
    archivePrefix = "arXiv",
    primaryClass = "astro-ph.IM",
    doi = "10.1088/0264-9381/33/3/035010",
    journal = "Class. Quant. Grav.",
    volume = "33",
    number = "3",
    pages = "035010",
    year = "2016"
}

@article{Hu:2017mde,
    author = "Hu, Wen-Rui and Wu, Yue-Liang",
    title = "{The Taiji Program in Space for gravitational wave physics and the nature of gravity}",
    doi = "10.1093/nsr/nwx116",
    journal = "Natl. Sci. Rev.",
    volume = "4",
    number = "5",
    pages = "685--686",
    year = "2017"
}

@article{Boer:2008ct,
    author = "Boer, Daniel and Boomsma, Jorn K.",
    title = "{Spontaneous CP-violation in the strong interaction at theta = pi}",
    eprint = "0806.1669",
    archivePrefix = "arXiv",
    primaryClass = "hep-ph",
    doi = "10.1103/PhysRevD.78.054027",
    journal = "Phys. Rev. D",
    volume = "78",
    pages = "054027",
    year = "2008"
}

@article{Boomsma:2008gf,
    author = "Boomsma, Jorn K. and Boer, Daniel",
    title = "{Spontaneous CP violation in the NJL model at theta = pi}",
    eprint = "0812.3077",
    archivePrefix = "arXiv",
    primaryClass = "hep-ph",
    doi = "10.1016/j.nuclphysa.2009.01.062",
    journal = "PoS",
    volume = "CONFINEMENT8",
    pages = "134",
    year = "2008"
}

@article{Kashiwa:2006rc,
    author = "Kashiwa, Kouji and Kouno, Hiroaki and Sakaguchi, Tomohiko and Matsuzaki, Masayuki and Yahiro, Masanobu",
    title = "{Chiral phase transition in an extended NJL model with higher-order multi-quark interactions}",
    eprint = "nucl-th/0608078",
    archivePrefix = "arXiv",
    reportNumber = "SAGA-HE-229-06",
    doi = "10.1016/j.physletb.2007.01.061",
    journal = "Phys. Lett. B",
    volume = "647",
    pages = "446--451",
    year = "2007"
}

@article{Osipov:2006ns,
    author = "Osipov, Alexander A. and Hiller, Brigitte and Blin, Alex H. and da Providencia, Joao",
    title = "{Effects of eight-quark interactions on the hadronic vacuum and mass spectra of light mesons}",
    eprint = "hep-ph/0607066",
    archivePrefix = "arXiv",
    doi = "10.1016/j.aop.2006.08.004",
    journal = "Annals Phys.",
    volume = "322",
    pages = "2021--2054",
    year = "2007"
}

@article{Osipov:2006ev,
    author = "Osipov, A. A. and Hiller, B. and Moreira, J. and Blin, A. H. and da Providencia, J.",
    title = "{Lowering the critical temperature with eight-quark interactions}",
    eprint = "hep-ph/0612082",
    archivePrefix = "arXiv",
    doi = "10.1016/j.physletb.2007.01.026",
    journal = "Phys. Lett. B",
    volume = "646",
    pages = "91--94",
    year = "2007"
}

@article{Asakawa:1989bq,
    author = "Asakawa, M. and Yazaki, K.",
    title = "{Chiral Restoration at Finite Density and Temperature}",
    doi = "10.1016/0375-9474(89)90002-X",
    journal = "Nucl. Phys. A",
    volume = "504",
    pages = "668--684",
    year = "1989"
}

@article{Helmboldt:2019pan,
    author = "Helmboldt, Alexander J. and Kubo, Jisuke and van der Woude, Susan",
    title = "{Observational prospects for gravitational waves from hidden or dark chiral phase transitions}",
    eprint = "1904.07891",
    archivePrefix = "arXiv",
    primaryClass = "hep-ph",
    doi = "10.1103/PhysRevD.100.055025",
    journal = "Phys. Rev. D",
    volume = "100",
    number = "5",
    pages = "055025",
    year = "2019"
}

@article{Corbin:2005ny,
    author = "Corbin, Vincent and Cornish, Neil J.",
    title = "{Detecting the cosmic gravitational wave background with the big bang observer}",
    eprint = "gr-qc/0512039",
    archivePrefix = "arXiv",
    doi = "10.1088/0264-9381/23/7/014",
    journal = "Class. Quant. Grav.",
    volume = "23",
    pages = "2435--2446",
    year = "2006"
}

@article{Crowder:2005nr,
    author = "Crowder, Jeff and Cornish, Neil J.",
    title = "{Beyond LISA: Exploring future gravitational wave missions}",
    eprint = "gr-qc/0506015",
    archivePrefix = "arXiv",
    doi = "10.1103/PhysRevD.72.083005",
    journal = "Phys. Rev. D",
    volume = "72",
    pages = "083005",
    year = "2005"
}

@article{Seto:2001qf,
    author = "Seto, Naoki and Kawamura, Seiji and Nakamura, Takashi",
    title = "{Possibility of direct measurement of the acceleration of the universe using 0.1-Hz band laser interferometer gravitational wave antenna in space}",
    eprint = "astro-ph/0108011",
    archivePrefix = "arXiv",
    doi = "10.1103/PhysRevLett.87.221103",
    journal = "Phys. Rev. Lett.",
    volume = "87",
    pages = "221103",
    year = "2001"
}

@article{Kawamura:2020pcg,
    author = "Kawamura, Seiji and others",
    title = "{Current status of space gravitational wave antenna DECIGO and B-DECIGO}",
    eprint = "2006.13545",
    archivePrefix = "arXiv",
    primaryClass = "gr-qc",
    doi = "10.1093/ptep/ptab019",
    journal = "PTEP",
    volume = "2021",
    number = "5",
    pages = "05A105",
    year = "2021"
}

@article{Kang:2025nhe,
    author = "Kang, Zhaofeng and Zhu, Jiang",
    title = "{Dark chiral phase transition driven by chemical potential and its gravitational wave test}",
    eprint = "2501.15242",
    archivePrefix = "arXiv",
    primaryClass = "hep-ph",
    doi = "10.1007/JHEP09(2025)005",
    journal = "JHEP",
    volume = "09",
    pages = "005",
    year = "2025"
}

@article{Garcia-Cely:2024ivo,
    author = "Garc{\'\i}a-Cely, Camilo and Landini, Giacomo and Zapata, {\'O}scar",
    title = "{Dark matter in QCD-like theories with a theta vacuum: Cosmological and astrophysical implications}",
    eprint = "2405.10367",
    archivePrefix = "arXiv",
    primaryClass = "hep-ph",
    doi = "10.1103/PhysRevD.111.063044",
    journal = "Phys. Rev. D",
    volume = "111",
    number = "6",
    pages = "063044",
    year = "2025"
}

@article{Klevansky:1992qe,
    author = "Klevansky, S. P.",
    title = "{The Nambu-Jona-Lasinio model of quantum chromodynamics}",
    doi = "10.1103/RevModPhys.64.649",
    journal = "Rev. Mod. Phys.",
    volume = "64",
    pages = "649--708",
    year = "1992"
}

@article{Hatsuda:1994pi,
    author = "Hatsuda, Tetsuo and Kunihiro, Teiji",
    title = "{QCD phenomenology based on a chiral effective Lagrangian}",
    eprint = "hep-ph/9401310",
    archivePrefix = "arXiv",
    reportNumber = "UTHEP-270, RYUTHP-94-1",
    doi = "10.1016/0370-1573(94)90022-1",
    journal = "Phys. Rept.",
    volume = "247",
    pages = "221--367",
    year = "1994"
}

@article{Stephanov:1998dy,
    author = "Stephanov, Misha A. and Rajagopal, K. and Shuryak, Edward V.",
    title = "{Signatures of the tricritical point in QCD}",
    eprint = "hep-ph/9806219",
    archivePrefix = "arXiv",
    reportNumber = "ITP-SB-98-39, MIT-CTP-2748, SUNY-NTG-98-17",
    doi = "10.1103/PhysRevLett.81.4816",
    journal = "Phys. Rev. Lett.",
    volume = "81",
    pages = "4816--4819",
    year = "1998"
}

@article{Allton:2002zi,
    author = "Allton, C. R. and Ejiri, S. and Hands, S. J. and Kaczmarek, O. and Karsch, F. and Laermann, E. and Schmidt, C. and Scorzato, L.",
    title = "{The QCD thermal phase transition in the presence of a small chemical potential}",
    eprint = "hep-lat/0204010",
    archivePrefix = "arXiv",
    reportNumber = "SWAT-02-335, NSF-ITP-02-26, BI-TP-2002-06",
    doi = "10.1103/PhysRevD.66.074507",
    journal = "Phys. Rev. D",
    volume = "66",
    pages = "074507",
    year = "2002"
}

@article{Stephanov:2004wx,
    author = "Stephanov, Mikhail A.",
    editor = "Muller, Berndt and Tan, C. I.",
    title = "{QCD Phase Diagram and the Critical Point}",
    eprint = "hep-ph/0402115",
    archivePrefix = "arXiv",
    doi = "10.1143/PTPS.153.139",
    journal = "Prog. Theor. Phys. Suppl.",
    volume = "153",
    pages = "139--156",
    year = "2004"
}

@article{Ratti:2005jh,
    author = "Ratti, Claudia and Thaler, Michael A. and Weise, Wolfram",
    title = "{Phases of QCD: Lattice thermodynamics and a field theoretical model}",
    eprint = "hep-ph/0506234",
    archivePrefix = "arXiv",
    doi = "10.1103/PhysRevD.73.014019",
    journal = "Phys. Rev. D",
    volume = "73",
    pages = "014019",
    year = "2006"
}

@article{Ratti:2006gh,
    author = "Ratti, Claudia and Thaler, Michael A. and Weise, Wolfram",
    title = "{Phase diagram and thermodynamics of the PNJL model}",
    eprint = "nucl-th/0604025",
    archivePrefix = "arXiv",
    month = "4",
    year = "2006"
}

@article{Zhang:2006gu,
    author = "Zhang, Zhao and Liu, Yu-Xin",
    title = "{Coupling of pion condensate, chiral condensate and Polyakov loop in an extended NJL model}",
    eprint = "hep-ph/0610221",
    archivePrefix = "arXiv",
    doi = "10.1103/PhysRevC.75.064910",
    journal = "Phys. Rev. C",
    volume = "75",
    pages = "064910",
    year = "2007"
}

@article{Ghosh:2006qh,
    author = "Ghosh, Sanjay K. and Mukherjee, Tamal K. and Mustafa, Munshi G. and Ray, Rajarshi",
    title = "{Susceptibilities and speed of sound from PNJL model}",
    eprint = "hep-ph/0603050",
    archivePrefix = "arXiv",
    reportNumber = "SINP-TNP-06-04",
    doi = "10.1103/PhysRevD.73.114007",
    journal = "Phys. Rev. D",
    volume = "73",
    pages = "114007",
    year = "2006"
}

@article{Gao:2021nwz,
    author = "Gao, Fei and Oldengott, Isabel M.",
    title = "{Cosmology Meets Functional QCD: First-Order Cosmic QCD Transition Induced by Large Lepton Asymmetries}",
    eprint = "2106.11991",
    archivePrefix = "arXiv",
    primaryClass = "hep-ph",
    reportNumber = "FTUV/21-33, IFIC/21-23",
    doi = "10.1103/PhysRevLett.128.131301",
    journal = "Phys. Rev. Lett.",
    volume = "128",
    number = "13",
    pages = "131301",
    year = "2022"
}

@article{tHooft:1976rip,
    author = "'t Hooft, Gerard",
    editor = "Shifman, Mikhail A.",
    title = "{Symmetry Breaking Through Bell-Jackiw Anomalies}",
    reportNumber = "PRINT-76-0254 (HARVARD)",
    doi = "10.1103/PhysRevLett.37.8",
    journal = "Phys. Rev. Lett.",
    volume = "37",
    pages = "8--11",
    year = "1976"
}

@article{Kobayashi:1970ji,
    author = "Kobayashi, M. and Maskawa, T.",
    title = "{Chiral symmetry and eta-x mixing}",
    doi = "10.1143/PTP.44.1422",
    journal = "Prog. Theor. Phys.",
    volume = "44",
    pages = "1422--1424",
    year = "1970"
}

@article{Moreira:2013ura,
    author = "Moreira, J. and Hiller, B. and Broniowski, W. and Osipov, A. A. and Blin, A. H.",
    title = "{Nonuniform phases in a three-flavor Nambu-Jona-Lasinio model}",
    eprint = "1312.4942",
    archivePrefix = "arXiv",
    primaryClass = "hep-ph",
    doi = "10.1103/PhysRevD.89.036009",
    journal = "Phys. Rev. D",
    volume = "89",
    number = "3",
    pages = "036009",
    year = "2014"
}

@inproceedings{Hiller:2008rz,
    author = "Hiller, B. and Osipov, A. A. and Moreira, J. and Blin, A. H.",
    title = "{Impact of eight quark interactions on chiral phase transitions II: Thermal effects}",
    booktitle = "{13th International Conference on Selected Problems of Modern Theoretical Physics (SPMTP 08): Dedicated to the 100th Anniversary of the Birth of D.I. Blokhintsev (1908-1979)}",
    eprint = "0809.2515",
    archivePrefix = "arXiv",
    primaryClass = "hep-ph",
    month = "9",
    year = "2008"
}

@article{Eguchi:1976iz,
    author = "Eguchi, Tohru",
    title = "{A New Approach to Collective Phenomena in Superconductivity Models}",
    reportNumber = "EFI 76/20-CHICAGO",
    doi = "10.1103/PhysRevD.14.2755",
    journal = "Phys. Rev. D",
    volume = "14",
    pages = "2755",
    year = "1976"
}

@article{Osipov:2005sp,
    author = "Osipov, Alexander A. and Hiller, Brigitte and Bernard, Veronique and Blin, Alex H.",
    title = "{Aspects of U(A)(1) breaking in the Nambu and Jona-Lasinio model}",
    eprint = "hep-ph/0507226",
    archivePrefix = "arXiv",
    doi = "10.1016/j.aop.2006.02.010",
    journal = "Annals Phys.",
    volume = "321",
    pages = "2504--2534",
    year = "2006"
}

@article{Coleman:1973jx,
    author = "Coleman, Sidney R. and Weinberg, Erick J.",
    title = "{Radiative Corrections as the Origin of Spontaneous Symmetry Breaking}",
    doi = "10.1103/PhysRevD.7.1888",
    journal = "Phys. Rev. D",
    volume = "7",
    pages = "1888--1910",
    year = "1973"
}

@article{Hubbard:1959ub,
    author = "Hubbard, J.",
    title = "{Calculation of partition functions}",
    doi = "10.1103/PhysRevLett.3.77",
    journal = "Phys. Rev. Lett.",
    volume = "3",
    pages = "77--80",
    year = "1959"
}

@article{Coleman:1977py,
    author = "Coleman, Sidney R.",
    title = "{The Fate of the False Vacuum. 1. Semiclassical Theory}",
    reportNumber = "HUTP-77-A004",
    doi = "10.1103/PhysRevD.15.2929",
    journal = "Phys. Rev. D",
    volume = "15",
    pages = "2929--2936",
    year = "1977",
    note = "[Erratum: Phys.Rev.D 16, 1248 (1977)]"
}

@article{Hindmarsh:2020hop,
    author = {Hindmarsh, Mark B. and L{\"u}ben, Marvin and Lumma, Johannes and Pauly, Martin},
    title = "{Phase transitions in the early universe}",
    eprint = "2008.09136",
    archivePrefix = "arXiv",
    primaryClass = "astro-ph.CO",
    reportNumber = "MPP-2020-163, HIP-2020-27/TH",
    doi = "10.21468/SciPostPhysLectNotes.24",
    journal = "SciPost Phys. Lect. Notes",
    volume = "24",
    pages = "1",
    year = "2021"
}

@book{Kolb:1990vq,
    author = "Kolb, Edward W. and Turner, Michael S.",
    title = "{The Early Universe}",
    reportNumber = "FERMILAB-BOOK-1990-01",
    doi = "10.1201/9780429492860",
    isbn = "978-0-429-49286-0, 978-0-201-62674-2",
    publisher = "Taylor and Francis",
    volume = "69",
    month = "5",
    year = "2019"
}

@article{Husdal:2016haj,
    author = "Husdal, Lars",
    title = "{On Effective Degrees of Freedom in the Early Universe}",
    eprint = "1609.04979",
    archivePrefix = "arXiv",
    primaryClass = "astro-ph.CO",
    doi = "10.3390/galaxies4040078",
    journal = "Galaxies",
    volume = "4",
    number = "4",
    pages = "78",
    year = "2016"
}

@article{Borsanyi:2016ksw,
    author = "Borsanyi, Sz. and others",
    title = "{Calculation of the axion mass based on high-temperature lattice quantum chromodynamics}",
    eprint = "1606.07494",
    archivePrefix = "arXiv",
    primaryClass = "hep-lat",
    reportNumber = "DESY-16-105",
    doi = "10.1038/nature20115",
    journal = "Nature",
    volume = "539",
    number = "7627",
    pages = "69--71",
    year = "2016"
}

@article{Kosowsky:1992vn,
    author = "Kosowsky, Arthur and Turner, Michael S.",
    title = "{Gravitational radiation from colliding vacuum bubbles: envelope approximation to many bubble collisions}",
    eprint = "astro-ph/9211004",
    archivePrefix = "arXiv",
    reportNumber = "FERMILAB-PUB-92-295-A",
    doi = "10.1103/PhysRevD.47.4372",
    journal = "Phys. Rev. D",
    volume = "47",
    pages = "4372--4391",
    year = "1993"
}

@article{Jinno:2016vai,
    author = "Jinno, Ryusuke and Takimoto, Masahiro",
    title = "{Gravitational waves from bubble collisions: An analytic derivation}",
    eprint = "1605.01403",
    archivePrefix = "arXiv",
    primaryClass = "astro-ph.CO",
    reportNumber = "KEK-TH-1900",
    doi = "10.1103/PhysRevD.95.024009",
    journal = "Phys. Rev. D",
    volume = "95",
    number = "2",
    pages = "024009",
    year = "2017"
}

@article{Jinno:2017ixd,
    author = "Jinno, Ryusuke and Lee, Sangjun and Seong, Hyeonseok and Takimoto, Masahiro",
    title = "{Gravitational waves from first-order phase transitions: Towards model separation by bubble nucleation rate}",
    eprint = "1708.01253",
    archivePrefix = "arXiv",
    primaryClass = "hep-ph",
    reportNumber = "CTPU-17-28, KEK-TH-1990",
    doi = "10.1088/1475-7516/2017/11/050",
    journal = "JCAP",
    volume = "11",
    pages = "050",
    year = "2017"
}

@article{Hindmarsh:2013xza,
    author = "Hindmarsh, Mark and Huber, Stephan J. and Rummukainen, Kari and Weir, David J.",
    title = "{Gravitational waves from the sound of a first order phase transition}",
    eprint = "1304.2433",
    archivePrefix = "arXiv",
    primaryClass = "hep-ph",
    reportNumber = "HIP-2013-07-TH",
    doi = "10.1103/PhysRevLett.112.041301",
    journal = "Phys. Rev. Lett.",
    volume = "112",
    pages = "041301",
    year = "2014"
}

@article{Hindmarsh:2015qta,
    author = "Hindmarsh, Mark and Huber, Stephan J. and Rummukainen, Kari and Weir, David J.",
    title = "{Numerical simulations of acoustically generated gravitational waves at a first order phase transition}",
    eprint = "1504.03291",
    archivePrefix = "arXiv",
    primaryClass = "astro-ph.CO",
    reportNumber = "HIP-2015-13-TH",
    doi = "10.1103/PhysRevD.92.123009",
    journal = "Phys. Rev. D",
    volume = "92",
    number = "12",
    pages = "123009",
    year = "2015"
}

@article{Kahniashvili:2009mf,
    author = "Kahniashvili, Tina and Kisslinger, Leonard and Stevens, Trevor",
    title = "{Gravitational Radiation Generated by Magnetic Fields in Cosmological Phase Transitions}",
    eprint = "0905.0643",
    archivePrefix = "arXiv",
    primaryClass = "astro-ph.CO",
    doi = "10.1103/PhysRevD.81.023004",
    journal = "Phys. Rev. D",
    volume = "81",
    pages = "023004",
    year = "2010"
}

@article{RoperPol:2019wvy,
    author = "Roper Pol, Alberto and Mandal, Sayan and Brandenburg, Axel and Kahniashvili, Tina and Kosowsky, Arthur",
    title = "{Numerical simulations of gravitational waves from early-universe turbulence}",
    eprint = "1903.08585",
    archivePrefix = "arXiv",
    primaryClass = "astro-ph.CO",
    reportNumber = "NORDITA-2019-024",
    doi = "10.1103/PhysRevD.102.083512",
    journal = "Phys. Rev. D",
    volume = "102",
    number = "8",
    pages = "083512",
    year = "2020"
}

@article{Ellis:2019oqb,
    author = "Ellis, John and Lewicki, Marek and No, Jos{\'e} Miguel and Vaskonen, Ville",
    title = "{Gravitational wave energy budget in strongly supercooled phase transitions}",
    eprint = "1903.09642",
    archivePrefix = "arXiv",
    primaryClass = "hep-ph",
    reportNumber = "KCL-PH-TH/2019-32, CERN-TH-2019-032, IFT-UAM/CSIC-19-32",
    doi = "10.1088/1475-7516/2019/06/024",
    journal = "JCAP",
    volume = "06",
    pages = "024",
    year = "2019"
}

@article{Christian:2025dhe,
    author = "Christian, Jan-Erik and Rather, Ishfaq Ahmad and Gholami, Hosein and Hofmann, Marco",
    title = "{Comprehensive analysis of constructing hybrid stars with a renormalization group-consistent Nambu-Jona-Lasino model}",
    eprint = "2503.13626",
    archivePrefix = "arXiv",
    primaryClass = "astro-ph.HE",
    doi = "10.1051/0004-6361/202555009",
    journal = "Astron. Astrophys.",
    volume = "701",
    pages = "A145",
    year = "2025"
}

@article{Buballa:1996tm,
    author = "Buballa, Michael",
    title = "{The Problem of matter stability in the Nambu-Jona-Lasinio model}",
    eprint = "nucl-th/9609044",
    archivePrefix = "arXiv",
    reportNumber = "SUNY-NTG-96-7",
    doi = "10.1016/S0375-9474(96)00314-4",
    journal = "Nucl. Phys. A",
    volume = "611",
    pages = "393--408",
    year = "1996"
}

@article{Costa:2008gr,
    author = "Costa, Pedro and de Sousa, C. A. and Ruivo, M. C. and Hansen, H.",
    title = "{The QCD critical end point in the PNJL model}",
    eprint = "0801.3616",
    archivePrefix = "arXiv",
    primaryClass = "hep-ph",
    doi = "10.1209/0295-5075/86/31001",
    journal = "EPL",
    volume = "86",
    number = "3",
    pages = "31001",
    year = "2009"
}

@article{Xia:2013caa,
    author = "Xia, Tao and He, Lianyi and Zhuang, Pengfei",
    title = "{Three-flavor Nambu{\textendash}Jona-Lasinio model at finite isospin chemical potential}",
    eprint = "1307.4622",
    archivePrefix = "arXiv",
    primaryClass = "hep-ph",
    doi = "10.1103/PhysRevD.88.056013",
    journal = "Phys. Rev. D",
    volume = "88",
    number = "5",
    pages = "056013",
    year = "2013"
}

@article{Sakai:2010rp,
    author = "Sakai, Yuji and Sasaki, Takahiro and Kouno, Hiroaki and Yahiro, Masanobu",
    title = "{Entanglement between deconfinement transition and chiral symmetry restoration}",
    eprint = "1006.3648",
    archivePrefix = "arXiv",
    primaryClass = "hep-ph",
    reportNumber = "SAGA-HE-257",
    doi = "10.1103/PhysRevD.82.076003",
    journal = "Phys. Rev. D",
    volume = "82",
    pages = "076003",
    year = "2010"
}

@article{Kashiwa:2007hw,
    author = "Kashiwa, Kouji and Kouno, Hiroaki and Matsuzaki, Masayuki and Yahiro, Masanobu",
    title = "{Critical endpoint in the Polyakov-loop extended NJL model}",
    eprint = "0710.2180",
    archivePrefix = "arXiv",
    primaryClass = "hep-ph",
    reportNumber = "SAGA-HE-237-07",
    doi = "10.1016/j.physletb.2008.01.075",
    journal = "Phys. Lett. B",
    volume = "662",
    pages = "26--32",
    year = "2008"
}

@article{Fukushima:2003fw,
    author = "Fukushima, Kenji",
    title = "{Chiral effective model with the Polyakov loop}",
    eprint = "hep-ph/0310121",
    archivePrefix = "arXiv",
    reportNumber = "MIT-CTP-3424",
    doi = "10.1016/j.physletb.2004.04.027",
    journal = "Phys. Lett. B",
    volume = "591",
    pages = "277--284",
    year = "2004"
}

@article{Schwaller:2015tja,
    author = "Schwaller, Pedro",
    title = "{Gravitational Waves from a Dark Phase Transition}",
    eprint = "1504.07263",
    archivePrefix = "arXiv",
    primaryClass = "hep-ph",
    reportNumber = "CERN-PH-TH-2015-093",
    doi = "10.1103/PhysRevLett.115.181101",
    journal = "Phys. Rev. Lett.",
    volume = "115",
    number = "18",
    pages = "181101",
    year = "2015"
}

@article{Zhao:2026pvt,
    author = "Zhao, Ruotong and Zhang, Zhao",
    title = "{Impact of chirality imbalance and nonlocal interactions on the QCD biased axionic domainwall interpretation of NANOGrav 15 year data}",
    eprint = "2603.07739",
    archivePrefix = "arXiv",
    primaryClass = "hep-ph",
    month = "3",
    year = "2026"
}

@article{Bertone:2004pz,
    author = "Bertone, Gianfranco and Hooper, Dan and Silk, Joseph",
    title = "{Particle dark matter: Evidence, candidates and constraints}",
    eprint = "hep-ph/0404175",
    archivePrefix = "arXiv",
    reportNumber = "FERMILAB-PUB-04-047-A",
    doi = "10.1016/j.physrep.2004.08.031",
    journal = "Phys. Rept.",
    volume = "405",
    pages = "279--390",
    year = "2005"
}

@article{Kribs:2016cew,
    author = "Kribs, Graham D. and Neil, Ethan T.",
    title = "{Review of strongly-coupled composite dark matter models and lattice simulations}",
    eprint = "1604.04627",
    archivePrefix = "arXiv",
    primaryClass = "hep-ph",
    doi = "10.1142/S0217751X16430041",
    journal = "Int. J. Mod. Phys. A",
    volume = "31",
    number = "22",
    pages = "1643004",
    year = "2016"
}

@article{Sakharov:1967dj,
    author = "Sakharov, A. D.",
    title = "{Violation of CP Invariance, C asymmetry, and baryon asymmetry of the universe}",
    doi = "10.1070/PU1991v034n05ABEH002497",
    journal = "Pisma Zh. Eksp. Teor. Fiz.",
    volume = "5",
    pages = "32--35",
    year = "1967"
}

@article{Morrissey:2012db,
    author = "Morrissey, David E. and Ramsey-Musolf, Michael J.",
    title = "{Electroweak baryogenesis}",
    eprint = "1206.2942",
    archivePrefix = "arXiv",
    primaryClass = "hep-ph",
    reportNumber = "NPAC-12-08",
    doi = "10.1088/1367-2630/14/12/125003",
    journal = "New J. Phys.",
    volume = "14",
    pages = "125003",
    year = "2012"
}

@article{Cohen:1993nk,
    author = "Cohen, Andrew G. and Kaplan, D. B. and Nelson, A. E.",
    title = "{Progress in electroweak baryogenesis}",
    eprint = "hep-ph/9302210",
    archivePrefix = "arXiv",
    reportNumber = "UCSD-PTH-93-02, BUHEP-93-4",
    doi = "10.1146/annurev.ns.43.120193.000331",
    journal = "Ann. Rev. Nucl. Part. Sci.",
    volume = "43",
    pages = "27--70",
    year = "1993"
}

@article{Cho:1992rv,
    author = "Cho, Peter L.",
    title = "{Chiral estimates of strong CP violation revisited}",
    eprint = "hep-ph/9212274",
    archivePrefix = "arXiv",
    reportNumber = "CALT-68-1842",
    doi = "10.1103/PhysRevD.48.3304",
    journal = "Phys. Rev. D",
    volume = "48",
    pages = "3304--3309",
    year = "1993"
}

@article{Crewther:1979pi,
    author = "Crewther, R. J. and Di Vecchia, P. and Veneziano, G. and Witten, Edward",
    title = "{Chiral Estimate of the Electric Dipole Moment of the Neutron in Quantum Chromodynamics}",
    reportNumber = "CERN-TH-2735",
    doi = "10.1016/0370-2693(79)90128-X",
    journal = "Phys. Lett. B",
    volume = "88",
    pages = "123",
    year = "1979",
    note = "[Erratum: Phys.Lett.B 91, 487 (1980)]"
}

@article{Pich:1991fq,
    author = "Pich, Antonio and de Rafael, Eduardo",
    title = "{Strong CP violation in an effective chiral Lagrangian approach}",
    reportNumber = "CERN-TH-6071-91",
    doi = "10.1016/0550-3213(91)90019-T",
    journal = "Nucl. Phys. B",
    volume = "367",
    pages = "313--333",
    year = "1991"
}

@article{Kibble:1976sj,
    author = "Kibble, T. W. B.",
    title = "{Topology of Cosmic Domains and Strings}",
    reportNumber = "ICTP/75/5",
    doi = "10.1088/0305-4470/9/8/029",
    journal = "J. Phys. A",
    volume = "9",
    pages = "1387--1398",
    year = "1976"
}

@article{Zeldovich:1974uw,
    author = "Zeldovich, Ya. B. and Kobzarev, I. Yu. and Okun, L. B.",
    title = "{Cosmological Consequences of the Spontaneous Breakdown of Discrete Symmetry}",
    reportNumber = "SLAC-TRANS-0165, IPM-MOSCOW-15",
    journal = "Zh. Eksp. Teor. Fiz.",
    volume = "67",
    pages = "3--11",
    year = "1974"
}

@article{Saikawa:2017hiv,
    author = "Saikawa, Ken'ichi",
    title = "{A review of gravitational waves from cosmic domain walls}",
    eprint = "1703.02576",
    archivePrefix = "arXiv",
    primaryClass = "hep-ph",
    reportNumber = "DESY-17-036",
    doi = "10.3390/universe3020040",
    journal = "Universe",
    volume = "3",
    number = "2",
    pages = "40",
    year = "2017"
}

@article{Hiramatsu:2013qaa,
    author = "Hiramatsu, Takashi and Kawasaki, Masahiro and Saikawa, Ken'ichi",
    title = "{On the estimation of gravitational wave spectrum from cosmic domain walls}",
    eprint = "1309.5001",
    archivePrefix = "arXiv",
    primaryClass = "astro-ph.CO",
    reportNumber = "ICRR-REPORT-659-2013-8, IPMU13-0182, YITP-13-87",
    doi = "10.1088/1475-7516/2014/02/031",
    journal = "JCAP",
    volume = "02",
    pages = "031",
    year = "2014"
}

@article{Hiramatsu:2010yz,
    author = "Hiramatsu, Takashi and Kawasaki, Masahiro and Saikawa, Ken'ichi",
    title = "{Gravitational Waves from Collapsing Domain Walls}",
    eprint = "1002.1555",
    archivePrefix = "arXiv",
    primaryClass = "astro-ph.CO",
    reportNumber = "ICRR-REPORT-559-2009-21, IPMU10-0024",
    doi = "10.1088/1475-7516/2010/05/032",
    journal = "JCAP",
    volume = "05",
    pages = "032",
    year = "2010"
}

@article{Gleiser:1998na,
    author = "Gleiser, Marcelo and Roberts, Ronald",
    title = "{Gravitational waves from collapsing vacuum domains}",
    eprint = "astro-ph/9807260",
    archivePrefix = "arXiv",
    reportNumber = "DART-HEP-98-03",
    doi = "10.1103/PhysRevLett.81.5497",
    journal = "Phys. Rev. Lett.",
    volume = "81",
    pages = "5497--5500",
    year = "1998"
}

@article{Guada:2020xnz,
    author = "Guada, Victor and Nemev{\v{s}}ek, Miha and Pintar, Matev{\v{z}}",
    title = "{FindBounce: Package for multi-field bounce actions}",
    eprint = "2002.00881",
    archivePrefix = "arXiv",
    primaryClass = "hep-ph",
    doi = "10.1016/j.cpc.2020.107480",
    journal = "Comput. Phys. Commun.",
    volume = "256",
    pages = "107480",
    year = "2020"
}

@Article{Nambu:1961tp,
  author  = {Nambu, Yoichiro and Jona-Lasinio, G.},
  title   = {{Dynamical Model of Elementary Particles Based on an Analogy with Superconductivity. 1.}},
  doi     = {10.1103/PhysRev.122.345},
  editor  = {Eguchi, T.},
  pages   = {345--358},
  volume  = {122},
  journal = {Phys. Rev.},
  year    = {1961},
}

@article{Sesana:2019vho,
	author = "Sesana, Alberto and others",
	title = "{Unveiling the gravitational universe at $\mu$-Hz frequencies}",
	eprint = "1908.11391",
	archivePrefix = "arXiv",
	primaryClass = "astro-ph.IM",
	doi = "10.1007/s10686-021-09709-9",
	journal = "Exper. Astron.",
	volume = "51",
	number = "3",
	pages = "1333--1383",
	year = "2021"
}

@article{Tanaka:2026geo,
	author = "Tanaka, Masanori and Wang, Jun-Chen and Zhang, Jing-Jun",
	title = "{Chiral phase transition with primordial black holes: Distinct phase structure and catalysis}",
	eprint = "2602.06661",
	archivePrefix = "arXiv",
	primaryClass = "hep-ph",
	month = "2",
	year = "2026"
}
	
\end{document}